\documentclass[11pt, twocolumn]{article}
\usepackage[a4paper,
            left=0.5in,
            right=0.5in,
            top=0.6in,
            bottom=0.8in,
            footskip=0.4in]{geometry}
\usepackage[utf8]{inputenc}

\usepackage{amsmath,amssymb,amsfonts,amsthm,color}

\usepackage{graphicx}
\usepackage{subfigure}

\usepackage[usenames]{xcolor}

\usepackage{url}
\usepackage{hyperref}
\hypersetup{
  colorlinks   = true, 
  urlcolor     = blue, 
  linkcolor    = black, 
  citecolor   = black 
}

\usepackage{lipsum}

\newcommand\blfootnote[1]{
  \begingroup
  \renewcommand\thefootnote{}\footnote{#1}
  \addtocounter{footnote}{-1}
  \endgroup
}

\usepackage[numbers,sort&compress]{natbib}
\let\OLDthebibliography\thebibliography
\renewcommand\thebibliography[1]{
  \OLDthebibliography{#1}
  \setlength{\parskip}{2pt}
  \setlength{\itemsep}{0pt plus 0.3ex}
}

\allowdisplaybreaks

\newcommand{\gev}{~\mathrm{GeV}}
\newcommand{\tev}{~\mathrm{TeV}}

\newcommand\refeq[1]{Eq.~(\ref{#1})}

\newcommand\refse[1]{Sect.~\ref{#1}}

\newcommand\citere[1]{Ref.~\cite{#1}}
\newcommand\citeres[1]{Refs.~\cite{#1}}

\def\reffi#1{\mbox{Fig.~\ref{#1}}}

\newcommand{\sigCMS}{2.9}
\newcommand{\muCMS}{0.33}
\newcommand{\dmuCMSpl}{0.19}
\newcommand{\dmuCMSmi}{0.12}

\newcommand\plane[2]{$(#1, #2)$ plane}

\newcommand{\SH}[1]{{\color{black}#1}}

\newcommand{\GW}[1]{{\color{black}#1}}

\usepackage{abstract}

\author{thomas.biekoetter@kit.edu}

\date{\today}

\usepackage{titlesec}
\titleformat{\section}
{\normalfont\bfseries}{\thesection}{1em}{}
\titleformat{\subsection}
{\normalfont\bfseries}{\thesubsection}{1em}{  }

\begin{document}

\def\thefootnote{\fnsymbol{footnote}}

\twocolumn[
\begin{@twocolumnfalse}
\begin{flushright}
\footnotesize
  KA-TP-03-2023, ~~ 
  DESY-23-033, ~~ 
  IFT--UAM/CSIC-23-028 ~~
\end{flushright}

\begin{center}
{\large
\textbf{
The CMS di-photon excess at 95 GeV 
in view of the LHC Run~2 results
}
}
\vspace{0.4cm}

Thomas Biekötter$^1$\footnote{thomas.biekoetter@desy.de},
Sven Heinemeyer$^2$\footnote{Sven.Heinemeyer@cern.ch} and
Georg Weiglein$^{3,4}$\footnote{georg.weiglein@desy.de}\\[0.2em]

{\small

  $^1${\textit{
   Institute for Theoretical Physics,
   Karlsruhe Institute of Technology,\\
   Wolfgang-Gaede-Str.~1, 76131 Karlsruhe, Germany
 }}
 
 $^2${\textit{
   Instituto de Física Teórica UAM-CSIC, Cantoblanco, 28049,
   Madrid, Spain
 }}
 
 $^3${\textit{
   Deutsches Elektronen-Synchrotron DESY,
     Notkestr.~85, 22607 Hamburg, Germany
  }}
  
 $^4${\textit{
   II. Institut für Theoretische Physik, Universität Hamburg,\\
    Luruper Chaussee 149, 22761 Hamburg, Germany\\[0.4em]
  }}
  
}

\begin{abstract}
The CMS collaboration has recently reported the
results of a low-mass Higgs-boson search in
the di-photon final state based on the full
Run 2 data set with refined analysis 
techniques. The new results 
show an excess
of events at a mass of about $95\gev$ with a local significance of
$\sigCMS\,\sigma$, confirming a previously reported excess at about the 
same mass and similar significance
based
on the first-year Run 2 plus Run 1 data. 
The observed excess is compatible
with the limits obtained in the corresponding
ATLAS searches.
In this work, we discuss the di-photon excess
and show that it can be interpreted as the
lightest Higgs boson in the Two-Higgs doublet
model that is extended by a complex singlet (S2HDM) of
Yukawa types~II and~IV. 
We show that the second-lightest Higgs
boson is in good
agreement with the current
LHC Higgs-boson measurements of the state at 125 GeV,
and that the full scalar sector is compatible with
all theoretical and experimental constraints.
Furthermore, we discuss the di-photon excess
in conjunction with an excess in the
$b \bar b$ final state observed at LEP and
an excess observed by CMS in the di-tau
final state, which were found at comparable
masses with local significances of about
$2\sigma$ and $3\sigma$, respectively.
We find that the $b \bar b$ excess can
be well described together with the
di-photon excess in both types of the S2HDM.
However, the di-tau excess can only be
accommodated at the level of $1\sigma$ in type IV.
We also comment on the compatibility with 
supersymmetric scenarios and other extended Higgs 
sectors, and
we discuss how the potential signal
can be further analyzed at the LHC and at
future $e^+e^-$ colliders.
\end{abstract}

\end{center}
\end{@twocolumnfalse}
]

\renewcommand{\thefootnote}{\arabic{footnote}}
\setcounter{footnote}{0}

\section{Introduction}
\label{sec:intro}

In
the year 2012 the ATLAS and CMS collaborations
discovered a new
particle~\cite{Aad:2012tfa,Chatrchyan:2012xdj}.
Within the current experimental and theoretical uncertainties the properties 
of the observed particle are
consistent with the predictions for 
the Higgs boson of the
Standard Model (SM) with a mass
of approximately
$125\gev$~\cite{CMS:2022dwd,ATLAS:2022vkf}, 
but they are also compatible with many scenarios 
of physics beyond the SM (BSM).  
While the minimal scalar sector of the SM features
only one physical scalar particle,
BSM physics often gives rise to
extended Higgs sectors in which additional
scalar particles are present.
Accordingly, one of the primary objectives of the
LHC is the search for additional Higgs bosons,
which is of crucial importance for exploring the
underlying physics of electroweak symmetry 
breaking.\blfootnote{$^*$thomas.biekoetter@kit.edu,
$^\dagger$sven.heinemeyer@cern.ch,
$^\ddagger$georg.weiglein@desy.de
}

Searches for Higgs bosons below
$125\gev$ have been performed at
LEP~\cite{Abbiendi:2002qp,Barate:2003sz,
Schael:2006cr},
the Tevatron~\cite{Group:2012zca} and the
LHC~\cite{CMS:2015ocq,CMS:2018cyk,
CMS:2018rmh,ATLAS:2018xad,CMS:2022goy,ATLAS:2022abz,
CMSnew}.
Among them, searches for di-photon resonances are
particularly intriguing, as this decay mode 
constituted one of the
two discovery channels of the
Higgs boson at $125\gev$.
CMS has performed searches for scalar di-photon
resonances at~$8\tev$ and $13\tev$.
Results based on the $8\tev$ data and the
first year of Run~2 data at $13\tev$,
corresponding to an integrated luminosity of
$19.7\,\mathrm{fb}^{-1}$ and $35.9\,\mathrm{fb}^{-1}$,
respectively,
showed a local excess of $2.8\,\sigma$
at $95.3 \gev$~\cite{CMS:2015ocq,CMS:2018cyk}.
This excess, which is present in both the $8\tev$
and the $13\tev$ data set, 
received considerable attention 
already soon after it was made public,
see e.g.~\citeres{Cao:2016uwt,
Fox:2017uwr,
Richard:2017kot,
Haisch:2017gql,
Biekotter:2017xmf,
Liu:2018xsw,
Domingo:2018uim,
Biekotter:2019kde,
Cline:2019okt,
Cao:2019ofo,
Aguilar-Saavedra:2020wrj}.

Recently, CMS published the result based on
their full Run~2 data set
and with substantially refined analysis
techniques. This new analysis
confirmed the excess of di-photon
events at about $95\gev$~\cite{CMSnew}.
By combining the data from the first,
second, and third years of Run~2,
which were collected at
$13\tev$ and correspond to integrated
luminosities of $36.3\,\mathrm{fb}^{-1}$,
$41.5\,\mathrm{fb}^{-1}$ and
$54.4\,\mathrm{fb}^{-1}$, respectively,
CMS finds an excess with a local
significance of $\sigCMS\,\sigma$ at a
mass of $95.4\gev$.
This ``di-photon excess'' can be
described by a scalar
resonance with a signal strength
of~\cite{CMSnewtalk}
\begin{equation}
\mu_{\gamma\gamma}^{\rm exp} =\frac{\sigma^{\rm exp} \left( gg \to \phi \to \gamma\gamma \right)}
         {\sigma^{\rm SM}\left( gg \to H \to \gamma\gamma \right)}
     = \muCMS^{+\dmuCMSpl}_{-\dmuCMSmi} \ .
\label{muCMS}
\end{equation}
Here $\sigma^{\rm SM}$ denotes the cross section
for a hypothetical SM
Higgs boson at the same mass.
In comparison to the previously reported results that 
were based just on the Run~1 and the first-year Run~2 
data~\cite{CMS:2018cyk},
the inclusion of the data collected
in the second and third year of Run~2 and
the refined analysis techniques 
yield a local significance of the excess that is almost 
unchanged, 
while the central value of the signal strength
$\mu_{\gamma\gamma}^{\rm exp}$
in \refeq{muCMS}
is substantially smaller than the
value $\mu_{\gamma\gamma}^{\rm exp} = 0.6 \pm 0.2$
extracted 
from the previous 
results~\cite{CMS:2018cyk}.

Regarding the interpretation of the new result from CMS 
it is important to note
that the updated analysis not only considered more
data, but in comparison to \citere{CMS:2018cyk}
it also improves the background suppression
of misidentified $Z \to e^+ e^-$
Drell-Yan events, and it includes
further event classes requiring the presence
of additional jets. Since 
a possible signal at about $95\gev$
giving rise to a relatively small number of events 
would occur 
on top of a fluctuating background,
one cannot necessarily rely on the the naive expectation
that the significance of 
an excess caused by a statistical fluctuation should be 
reduced by the inclusion of more data while it should be 
increased in case of an actual signal.
In fact, even in the latter case 
the excess of events observed in the different
data sets and evaluated at a fixed mass value
would still be expected to fluctuate.
From our point of view the fact that the inclusion of 
the additional data sets and the improvements in the 
analysis have led to an excess of events at approximately 
the same mass as previously reported with a statistical 
significance that has not been reduced strengthens the 
motivation for exploring a possible BSM origin of the 
observed results.

ATLAS reported results of searches
in the di-photon final state below $125\gev$
using $80~\mathrm{fb}^{-1}$ of Run~2 data
in 2018~\cite{ATLAS:2018xad}.
The ATLAS search 
found only a mild excess of about $1\,\sigma$
local significance
at masses around $95\gev$.
However, the cross section limits
obtained in the ATLAS analysis
are substantially weaker than the corresponding
CMS limits, even in the mass range where
CMS reported the excess~\cite{Heinemeyer:2018wzl},
and the excess observed in CMS is therefore compatible 
with the ATLAS limits.

If the origin of the di-photon excesses
at $95\gev$ is a new particle, the question
arises whether it is also detectable in
other collider channels, and whether additional
indications for this new particle 
might have already occurred
in other existing searches.
Notably, LEP reported a local $2.3\,\sigma$ excess
in the~$e^+e^-\to Z(H\to b\bar{b})$
searches\,\cite{Barate:2003sz}, which would
be consistent with a
scalar particle with a mass of about 
$95\gev$.\footnote{Due to
the $b \bar b$ final state the 
mass resolution is significantly worse
compared to the resolution of
searches in the di-photon final state.}
This ``$b \bar b$ excess'' corresponds to
a signal strength of
$\mu_{bb}^{\rm exp} =
0.117 \pm 0.057$~\cite{Cao:2016uwt,Azatov:2012bz}.
Moreover, CMS observed another excess
compatible with a mass of $95\gev$ in
the Higgs-boson searches utilizing
di-tau final states~\cite{CMS:2022goy}.
This excess was
most pronounced at a mass of $100\gev$
with a local significance of $3.1\,\sigma$,
but it is also well compatible with a mass
of $95\gev$, where the local significance
amounts to $2.6\,\sigma$. For this
``di-tau excess'', the best-fit
signal strength for a mass hypothesis
of $95\gev$ was determined to be
$\mu^{\rm exp}_{\tau\tau} =
1.2 \pm 0.5$.
It is noteworthy that, to date,
ATLAS has not published a search in
the di-tau final state that covers the
mass range around 95~GeV.

Given that the excesses observed by CMS
and LEP occurred at a similar mass,
the intriguing question arises whether
the excesses in the three different channels might arise from the 
production of a single new particle.
This triggered activities in the literature regarding possible 
model interpretations that could account 
for the various excesses
while also satisfying all other
measurements related to the Higgs sector.
Models in which the previously observed
two excesses in the di-photon and the $b \bar b$
final states can be described simultaneously
(with the CMS excess based only on the Run~1 and
first year Run~2 data) were reviewed
in~\citeres{Heinemeyer:2018jcd,Heinemeyer:2018wzl}.
In \citere{Biekotter:2019kde}
those two excesses were studied in the \GW{Two-Higgs doublet model (2HDM)} with
an additional real singlet (N2HDM), with several follow-up
analyses~\cite{Biekotter:2021ovi,
Biekotter:2021qbc,Heinemeyer:2021msz},
while in \citeres{Biekotter:2022jyr,Biekotter:2022abc}
also the more recently observed excess
in the di-tau searches was taken into account.

\medskip
Since the new result obtained by CMS
confirmed the previously observed
di-photon excess
at about $95\gev$ 
but resulted in a significant change in
the required signal rate
$\mu_{\gamma\gamma}^{\rm exp}$, 
it is of interest to assess the implications of the new result on possible
model interpretations. In the present paper we focus in particular on 
the extension of the 2HDM by a complex singlet
(S2HDM) as a template for a model where 
a mostly gauge-singlet scalar particle
obtains its couplings to fermions
and gauge bosons via the mixing with
the SM-like Higgs boson at $125\gev$.
We will demonstrate that 
this kind of scenario is suitable for describing
the di-photon excess. 
In this context we will in particular investigate the impact of the 
reduced central value of the signal strength of
$\mu_{\gamma\gamma}^{\rm exp} = \muCMS$~\cite{CMSnew}
compared to the result of $\mu_{\gamma\gamma}^{\rm exp} = 0.6$
that was obtained based on the previous analysis~\cite{CMS:2018cyk}.
Moreover, we will discuss the possibility
of simultaneously
describing the $b\bar b$ excess
and the di-tau excess.
We will further discuss
possible ways in which the presented scenario
could be confirmed or excluded
experimentally in
the near future.

The paper is structured as follows. In \refse{sec:modeldef} 
we introduce the S2HDM and 
define our notation.
In \refse{sec:quanti}
we qualitatively discuss how
sizable signal rates
in the three channels in which the excesses
have been observed can arise.
The relevant theoretical and experimental
constraints on the model parameters
are discussed in \refse{sec:constraints}.
We present our numerical results and
discuss their implications in \refse{sec:num}, including
an analysis of future experimental prospects.
The conclusions and an outlook are given
in \refse{sec:conclu}.

\section{A 95~GeV Higgs boson in the S2HDM}

In this section we briefly summarize the
scalar sector of
S2HDM and how the excesses at $95\gev$
can be accommodated in this model.
We also discuss the relevant experimental
and theoretical constraints that we
apply in our numerical analysis.

\subsection{Model definitions}
\label{sec:modeldef}
In the SM the Higgs sector
contains a single SU(2) doublet $\Phi_1$.
The S2HDM extends the SM by a second
Higgs doublet field $\Phi_2$
and an additional
complex gauge-singlet field
$\Phi_S$~\cite{Jiang:2019soj,
Biekotter:2021ovi}.
The richer structure of the scalar sector is
motivated for instance by the possibility
of a first-order electroweak phase
transition~\cite{Biekotter:2021ysx},
and the related phenomenology, including
electroweak baryogenesis, or the
presence of a stochastic primordial
gravitational-wave
background.
From a more theoretical perspective,
the presence of a second
Higgs doublet field arises in several
extensions of the SM that address the
hierarchy problem in the context of
supersymmetry~\cite{Fayet:1976et}
or compositeness~\cite{Mrazek:2011iu},
and in many models addressing the
strong CP problem of QCD~\cite{Kim:1986ax}.
Due to the presence of the complex scalar
singlet field, the S2HDM can accommodate
a dark-matter candidate in the form of
pseudo-Nambu-Goldstone (pNG) dark
matter~\cite{Barger:2008jx}.
As will be discussed below,
among the various
proposed WIMP dark-matter candidates,
pNG dark matter is 
in particular motivated in view of the
existing limits from
dark-matter
direct-detection
experiments~\cite{PandaX-4T:2021bab,
XENON:2018voc,LZ:2022ufs}.

The vacuum state of the S2HDM is characterized
by non-zero vacuum expectation values (vev)
$v_1$ and $v_2$
for the neutral CP-even components of the
Higgs doublets fields $\Phi_1$
and $\Phi_2$, respectively.
The presence of these vevs
leads to the spontaneous breaking of the
electroweak symmetry. As in the usual 2HDM,
one defines the parameter $\tan\beta = v_2 / v_1$,
where $v_1^2 + v_2^2 = v^2 \approx (246 \gev)^2$
corresponds to the SM vev squared.
In addition, the real component of the
singlet field has the non-zero vev $v_S$,
which breaks a global U(1) symmetry under which
only $\Phi_S$ is charged.
If this symmetry 
was exact initially,
the imaginary component of $\Phi_S$
would act as a massless Goldstone boson.
Therefore, one introduces a soft breaking
via a bilinear term
$-m_\chi^2 (\Phi_S^2 + \mathrm{h.c.})$,
which gives rise to a mass $m_\chi$
for the imaginary component of $\Phi_S$,
which then
plays the role of the pNG
dark-matter state.

Neglecting possible sources of CP violation,
as we do throughout this paper,
the physical scalar spectrum of the S2HDM consists
of three CP-even Higgs bosons $h_{1,2,3}$
with masses $m_{h_{1,2,3}}$ that
are mixed states composed of the neutral real
components of $\Phi_{1,2}$ and the real
component of $\Phi_S$. The imaginary
component of $\Phi_S$ does not mix 
with other states and results in
a stable scalar dark-matter particle which is labeled
$\chi$ in the following. Moreover, as in the
CP-conserving 2HDM, the scalar spectrum contains
a pair of charged Higgs bosons $H^\pm$ and
a CP-odd Higgs boson $A$ with masses
$m_{H^\pm}$ and $m_A$, respectively.

For the presence of two Higgs doublets,
the most general
gauge invariant Yukawa sector gives rise
to flavour-changing neutral currents (FCNC) at
the tree-level. These are, however, strongly
constrained experimentally.
In order to avoid FCNC at the tree-level,
we impose an additional 
$Z_2$ symmetry under which
one of the doublet fields changes the sign,
which is only softly-broken via a term of
the form $-m_{12}^2(\Phi_1^\dagger \Phi_2
+ \mathrm{h.c.})$.
This symmetry can be extended to the fermion
sector such that either $\Phi_1$ or $\Phi_2$
(but not both) couples
to either the charged leptons $\ell$, the up-type
quarks $u$ or the down-type quarks $d$.
There are four different possibilities to
assign conserved charges for the three kinds
of fermions, giving rise to the four Yukawa
types~I, II, III (lepton-specific) and
IV (flipped) that are known from the
$Z_2$-symmetric 2HDM
(see e.g.~\citere{Branco:2011iw}).

For the Yukawa types~II and~IV, $\Phi_1$
is coupled to down-type quarks
and $\Phi_2$ is coupled to up-type
quarks. In this case
an independent modification of the couplings
of the Higgs bosons $h_i$ to bottom quarks and
top quarks is possible.
These two types are therefore of particular interest 
regarding the prediction of a sufficiently large 
di-photon signal rate~\cite{Biekotter:2019kde}.

\subsection{Interpretation of the excesses}
\label{sec:quanti}
In the following discussion,
the lightest of the three CP-even Higgs
bosons of the S2HDM $h_1$
serves as the possible particle state
at $95\gev$, also denoted $h_{95}$ from here on.
We furthermore assume that
the second lightest Higgs boson, $h_2 = h_{125}$,
corresponds to the state discovered
at about $125 \gev$.
The key aspect of the signal interpretation
presented here is that $h_{95}$ obtains
its couplings to the
fermions and gauge
bosons as a result of the mixing with the
CP-even components of the two doublets.
In order to comply
with the constraints from the
Higgs-boson searches at LEP 
\GW{in the mass region of about $95\gev$}
and the LHC cross section measurements 
\GW{for the detected state at $125\gev$, 
the state} $h_{95}$ must have couplings
to gauge-bosons that are reduced by roughly
one order of magnitude as compared to the
couplings of a SM Higgs boson \GW{of the same mass}. 
As a consequence,
in the S2HDM interpretation $h_{95}$ is dominantly
singlet-like.

Despite the \GW{predominant singlet-like character} of $h_{95}$,
sizable decay rates into di-photon pairs can
be achieved via a suppression of the otherwise
dominating decay into $b$-quark 
pairs (see also \citere{Barbieri:2013nka}).
At the same time, 
\GW{no such suppression should occur for
the coupling} to top quarks,
whose loop contribution gives rise to the
decay into photons 
\GW{(and also governs the production process via gluon fusion)}.
As a result, large signal rates $\mu_{\gamma\gamma}$
can \GW{occur} in the S2HDM if
$|c_{h_{95} t \bar t} / c_{h_{95} b \bar b}| > 1$,
where the coupling coefficients $c_{h_{95} t \bar t}$
and $c_{h_{95} b \bar b}$ are the couplings
of $h_{95}$ to the respective quark normalized
to the couplings of a hypothetical
SM Higgs boson of the same mass.
It becomes apparent that
the Yukawa types~I and~III, 
\GW{for which}
$c_{h_{95} t \bar t} = c_{h_{95} b \bar b}$ 
\GW{applies,
do not feature the conditions for a sufficiently large
di-photon branching ratio in accordance with}
the CMS excess.
On the other hand, in type~II and type~IV
\GW{the two} coupling coefficients can be modified
independently. This can potentially
enhance the di-photon
branching ratio by up to
an order of magnitude~\cite{Biekotter:2019kde,
Biekotter:2022jyr}, such that
sizable values of $\mu_{\gamma\gamma}$
can be accommodated even for a relatively small
mixing with the detected Higgs boson
at $125\gev$ (and thus suppressed
cross sections).\footnote{An
additional, although not as significant,
enhancement of $\mu_{\gamma\gamma}$ can
\GW{occur} if $c_{h_{95} t \bar t}$ and
$c_{h_{95} b \bar b}$ carry a relative minus
sign. This relative sign gives \GW{rise to} constructive interference
effects in the loop-induced couplings of
$h_{95}$ to gluons and photons, hence enhancing
both the production and the decay rate in
the $gg \to h_{95} \to \gamma\gamma$ channel.}

Since larger values of $\mu_{\gamma\gamma}$
can be achieved in type~II and~IV
compared to type~I and type~III
as discussed above,
we will focus on the
type~II and the type~IV
in the following.
Between these two types, an important difference
arises from the fact that 
$c_{h_{95} \tau^+ \tau^-} =
c_{h_{95} b \bar b}$ 
\GW{holds}
in type~II, whereas
\GW{the corresponding relation in type~IV is}
$c_{h_{95} \tau^+ \tau^-} =
c_{h_{95} t \bar t}$.
Accordingly, 
\GW{in the parameter regions of
type~II where the di-photon signal rate is
enhanced as a consequence of the suppression of its coupling to
$b$-quark pairs}
the coupling of $h_{95}$
to tau-leptons is \GW{simultaneously}
suppressed.
Hence, type~II is not expected to yield
sizable signal rates in the $\tau^+ \tau^-$
decay channel if the di-photon excess
is accommodated.
On the other hand, given that
$c_{h_{95} t \bar t}$ 
\GW{should be}
unsuppressed for a description of
the di-photon excess,
\GW{type~IV can give rise to a simultaneous description of 
the CMS di-tau excess}~\cite{Biekotter:2022jyr}.

\subsection{Constraints}
\label{sec:constraints}

The parameter space that is relevant for
a possible description of the excesses at
$95\gev$ is subject to various theoretical
and experimental constraints. We will briefly
discuss the relevant constraints in the following.

Theoretical constraints that we \GW{apply in our analysis}
ensure that the perturbative treatment of
the scalar sector of the S2HDM is valid.
To this end, we demand
that the eigenvalues of the scalar
$2\times 2$ scattering
matrix in the high-energy limit
are smaller than $8\pi$, giving
rise to the so-called tree-level perturbative
unitarity constraints~\cite{Biekotter:2021ovi}.
In addition, using the approach
described in \citere{Biekotter:2021ovi}
we 
\GW{apply a condition on the stability of} 
the electroweak vacuum (\GW{see}
\refse{sec:modeldef}) 
by requiring that
the tree-level scalar potential
is bounded from below, and that the
electroweak vacuum corresponds to the
global minimum of the
potential.

Moreover, the parameters of the
S2HDM are constrained by various
experimental 
\GW{results.}
With regards to the collider phenomenology,
we check whether the parameter points
are in agreement with the cross section
limits from collider searches for
BSM Higgs bosons by making use of the
public code
\texttt{HiggsBounds v.6}~\cite{Bechtle:2008jh,
Bechtle:2011sb,
Bechtle:2013wla,
Bechtle:2020pkv,
Bahl:2022igd} (as part of the new code 
\texttt{HiggsTools}~\cite{Bahl:2022igd}).
A parameter point is rejected if
\GW{the} signal rate of one of the Higgs bosons
in the most sensitive search channel
(based on the expected limits) is larger
than the experimentally observed
limit at the 95\% confidence level.

In order to ensure that 
\GW{the properties of $h_{125}$ are}
in agreement
with the measured signal rates from the LHC,
we make use of the public code
\texttt{HiggsSignals v.3}~\cite{Bechtle:2013xfa,
Bechtle:2014ewa,
Bechtle:2020uwn,
Bahl:2022igd} (as part of the new code 
\texttt{HiggsTools}~\cite{Bahl:2022igd}).
This code performs a $\chi^2$ fit to
a large dataset of LHC cross section measurements
in the different channels in which the
SM-like Higgs boson was observed.
As a requirement for accepting or rejecting
a parameter point, we use the condition
$\chi^2_{125} \leq \chi^2_{{\rm SM},125} + 6.18$,
where $\chi^2_{125}$ is the fit value of
the S2HDM parameter point under consideration,
and $\chi^2_{{\rm SM},125} = 146.15$ is the fit result
assuming a Higgs boson at $125\gev$ that behaves
according to the predictions of the SM.
In two-dimensional parameter 
\GW{planes}
the above condition ensures that the 
\GW{selected}
S2HDM parameter points are not disfavoured
by more than $2\,\sigma$ in comparison to the SM
\GW{regarding the properties of $h_{125}$}.

Both \texttt{HiggsBounds} and \texttt{HiggsSignals}
require as input the cross sections and the
branching ratios of the scalar state 
\GW{for the considered parameter point}.
The cross sections were derived internally
in \texttt{HiggsBounds} from the
effective couplings coefficients.
For the computation of the branching ratios,
we applied the
library \texttt{N2HDECAY}~\cite{Muhlleitner:2016mzt,
Engeln:2018mbg},
which we
modified to account for decays
of the Higgs bosons into pairs of
the DM state $\chi$~\cite{Biekotter:2021ovi}.

Indirect experimental constraints on the
Higgs sector can be obtained from flavour-physics
observables and \GW{from} electroweak precision
observables.
Lacking precise theoretical predictions
for the different flavour observables in
the S2HDM,
we apply conservative
lower limits of $\tan\beta > 1.5$ and $m_{H^\pm} > 600\gev$
in our S2HDM parameter scans in type~II and type~IV
to \GW{ensure}
agreement with
the flavour-physics constraints~\cite{Haller:2018nnx}.
With regards to the electroweak precision observables,
we apply constraints in terms of the oblique parameters
$S$, $T$ and $U$ which we computed according to
\citere{Grimus:2008nb} at the one-loop level.
We required that the predicted values of the
oblique parameters are in agreement with the
fit result of \citere{Haller:2018nnx}
within a confidence level
of $2\,\sigma$.\footnote{The fit result of the oblique
parameters was obtained before the recent
CDF measurement of $M_W$~\cite{CDF:2022hxs},
which showed a significant
upward deviation with respect to the SM prediciton.
We demonstrated in \citere{Biekotter:2022abc}
that a larger value for the $W$-boson
mass, even as large as the
central value of the
CDF measurement, can
be accommodated in a 2HDM that is 
extended by a singlet if there
are sizable mass splittings between the heavy
BSM Higgs bosons $h_3$, $A$ and $H^\pm$,
while in addition
the excesses at $95\gev$ can be accommodated
in the same way as presented here.}

As a consequence of the presence of
the stable scalar state
$\chi$, further constraints on the S2HDM
\GW{parameter space} arise from the measurements of the
dark-matter relic abundance of the universe.
Assuming the freezout mechanism for the
production of $\chi$ in the early universe,
we applied the Planck measurement of today's
relic abundance
of $h^2 \Omega = 0.119$~\cite{Planck:2018vyg}
as an upper limit, thus avoiding overproduction
of dark matter. The theoretical predictions for
the relic abundance of $\chi$ were obtained by
making use of the public code
\texttt{micrOMEGAs}~\cite{Belanger:2018ccd}.

Given its nature 
\GW{as}
a pNG
boson of the softly-broken global U(1) symmetry,
the cross sections for the scattering of
$\chi$ on nuclei are highly suppressed in the
limit of small momentum transfer as relevant
for dark-matter direct detection
experiments~\cite{Barger:2008jx}.
As a result,
it has been shown that even including
loop corrections the current
direct detection
constraints are 
\GW{of minor importance}
in the
S2HDM~\cite{Biekotter:2022bxp}.
We nevertheless applied the
currently strongest
spin-independent
cross section limits for the scattering of
$\chi$ on nucleons obtained by the LZ
collaboration~\cite{LZ:2022ufs},
where we \GW{used} the one-loop predictions
of the scattering cross sections as computed
in \citere{Biekotter:2022bxp}.\footnote{Dark-matter
indirect detection experiments can so far only
probe a very limited mass window of $m_\chi$
once the experimental upper limit on the
relic abundance is applied~\cite{Biekotter:2021ovi}.
Thus, we do not
consider additional constraints from
indirect-detection experiments.}

We finally note that the DM constraints that
are imposed in our analysis could also be evaded
entirely assuming that the U(1) symmetry acting
on~$\Phi_S$ is gauged~\cite{Gross:2017dan}.
In this case the imaginary
component of~$\Phi_S$ in general is not stable.
In an effective field theory framework, the
decay is described by higher-dimensional operators
that are suppressed by powers of the U(1)-breaking
scale. Depending on the size of this scale, the
lifetime of~$\chi$ could be comparable or larger
than the age of the universe, in which case $\chi$ can
still be a viable candidate for (decaying)~DM,
or $\chi$ could be short-lived and thus would not contribute to
the~DM relic abundance. In the latter case,
the constraints from the measured DM relic abundance
and DM direct detection experiments do not
apply, but on the other hand in this case the model looses the
attractive feature of providing a~pNG DM state.
The most studied model realizations of this kind assume
that the U(1) corresponds to a gauged
U(1)$_L$~\cite{Abe:2020dut} or
U(1)$_{B - L}$~\cite{Abe:2020iph,Okada:2020zxo,
Mohapatra:2023aei} symmetry,
where~$L$ and~$B$ stand for lepton number
and baryon number, respectively,
such that $\Phi_S$ carries lepton number
and can in particular decay into neutrinos.
Another possibility is a hidden
U(1)$_D$ symmetry in the dark sector, where
the kinetic mixing between the U(1)$_D$ and
U(1)$_Y$ gauge fields is responsible for
the decay of~$\chi$~\cite{Liu:2022evb}.
In any case, our conclusions regarding the
description of the excesses at~95~GeV
do not rely on the application of the 
DM constraints, see also the discussion below.

\section{Numerical discussion}
\label{sec:num}

In order to address the question whether
a description of the CMS di-photon excess 
can be realized in the S2HDM,
possibly in combination with
the excesses in the $b \bar b$ and the
di-tau final states, we performed a
parameter scan in the Yukawa types~II
and~IV of the S2HDM.
\GW{We investigated the theoretical predictions in comparison to 
the experimental results for the observed excesses near $95\gev$, 
ensuring at the same time that the properties of the}  
Higgs boson at $125\gev$ 
\GW{are}
in good agreement with the most
up-to-date LHC signal rate measurements.
To this end, we implemented a
genetic algorithm (using the python
package \texttt{DEAP}~\cite{DEAP_JMLR2012})
that minimizes
a loss function constructed 
\GW{from}
$\chi^2_{125}$ (obtained using
\texttt{HiggsSignals}) and the three
contributions $\chi^2_{\gamma\gamma}$,
$\chi^2_{bb}$,
and $\chi^2_{\tau\tau}$
quantifying the
\GW{compatibility with}
the excesses at $95\gev$,
where we define the latter as
\begin{equation}
\chi^2_{\gamma\gamma,\tau\tau,bb} =
\frac{
(\mu_{\gamma\gamma,\tau\tau,bb} -
\mu_{\gamma\gamma,\tau\tau,bb}^{\rm exp})^2 }{
(\Delta \mu_{\gamma\gamma,\tau\tau,bb}^{\rm exp})^2} \ .
\label{eq:chisq95indi}
\end{equation}
Here the experimental central
values and the uncertainties were
stated in \refse{sec:intro}, and
$\mu_{\gamma\gamma,\tau\tau,bb}$
are the theoretically predicted values.
Since $\mu_{\gamma\gamma}^{\rm exp}$ has
asymmetric uncertainties, we define
$\chi^2_{\gamma\gamma}$ in such a way that
the lower uncertainty is used if
$\mu_{\gamma\gamma} < \mu_{\gamma\gamma}^{\rm exp}$,
and the upper uncertainty
is used if
$\mu_{\gamma\gamma} > \mu_{\gamma\gamma}^{\rm exp}$.
To obtain the predictions for
$\mu_{\gamma\gamma}$ and
$\mu_{\tau\tau}$,
we used \texttt{HiggsTools}
to derive the gluon-fusion cross section
of the state at $95\gev$
via a re-scaling of the SM predictions
as a function of $c_{h_{95} t \bar t}$ and
$c_{h_{95} b \bar b}$. To compute
$\mu_{b b}$, we \GW{approximated}
the cross section
ratio as $\sigma / \sigma_{\rm SM}
= c_{h_{95} VV}^2$.
The branching ratios
of $h_{95}$ were obtained with the help
of \texttt{N2HDECAY} (see
also the discussion in \refse{sec:constraints}).

The set of parameter points obtained by
the minimization of
the loss function
was then confronted with
the constraints discussed in \refse{sec:constraints}.
\GW{Parameter points 
that did not pass the applied constraints were rejected.}
For the generation of the S2HDM parameter
points and the application of the
constraints, we used the program
\texttt{s2hdmTools}~\cite{Biekotter:2021ovi,
Biekotter:2022bxp}, which features interfaces
to \texttt{HiggsBounds}, \texttt{HiggsSignals},
\texttt{micrOMEGAs} and \texttt{N2HDECAY}.

We chose the values of the free parameters in
our scan as follows. The mass of $h_{95}$
was varied in the region in which the
di-photon excess is most pronounced,
i.e.~$94\gev \leq m_{h_{95}} \leq 97\gev$.
The mass of the second-lightest Higgs boson
\GW{was} set to $m_{h_{125}} = 125.09\gev$, and
the third heavier Higgs boson, denoted $H$
in the following, was scanned freely up to an
upper limit of \GW{$m_H = 1\tev$}.
The same upper limit was chosen for the masses
of the DM state $\chi$, the 
\GW{CP-odd}
Higgs boson $A$, and the charged Higgs bosons
$H^\pm$, where for the latter additionally
the lower limit $m_{H^\pm} > 600\gev$ was applied
\GW{arising from}
the flavour constraints.
Moreover, we varied $\tan\beta$ in the
range $1.5 \leq \tan\beta\leq 10$, and for the
singlet vev we chose $40\gev \leq v_S \leq 2\tev$.
Finally, the scan range of the parameter
$m_{12}^2$ was determined by the condition
$400\gev \leq M \leq 1\tev$, where
$M^2 = m_{12}^2 / (\sin\beta \cos\beta)$.

\subsection{Description of the
di-photon excess}

\begin{figure}[t]
\centering
\includegraphics[width=0.9\columnwidth]{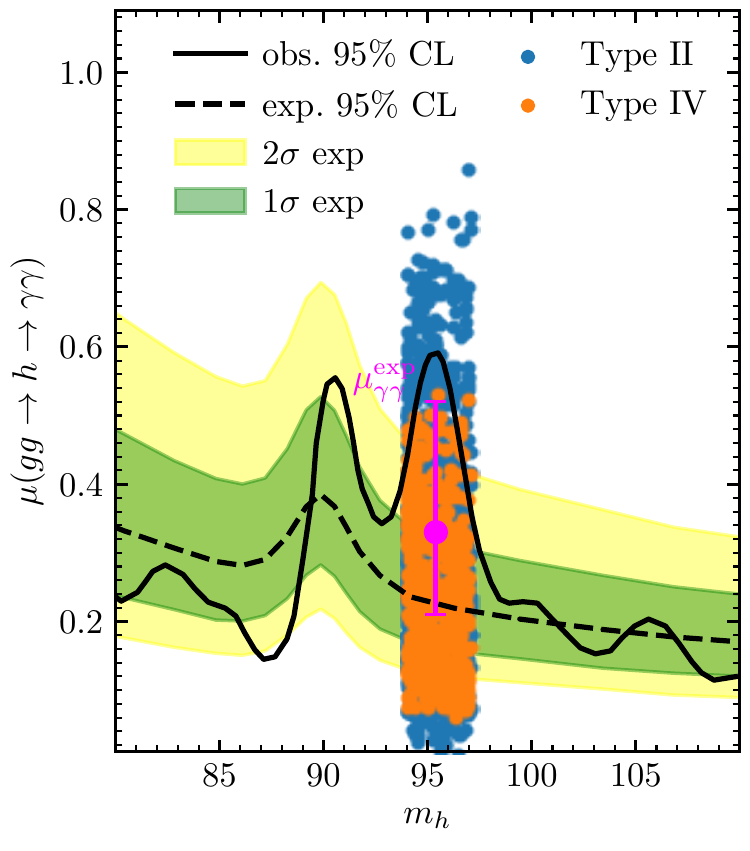}~
\vspace*{-0.4cm}
\caption{\small
S2HDM parameter points
\GW{passing the applied constraints}
in the \plane{m_{h_{95}}}{\mu_{\gamma\gamma}}
for the type~II (blue) and the type~IV (orange).
The expected and observed
cross section limits 
\GW{obtained}
by CMS are indicated
\GW{by}
the black dashed and solid lines, respectively,
and the $1\sigma$ and $2\sigma$ uncertainty intervals
are indicated 
\GW{by}
the green and yellow
bands, respectively. The value of
$\mu_{\gamma\gamma}^{\rm exp}$ and its uncertainty
is shown with the magenta error bar at the mass
\GW{value}
at which the excess is most pronounced.}
\label{fig:gamgam}
\end{figure}

In \reffi{fig:gamgam} we show the
predictions for $\mu_{\gamma\gamma}$ for the
S2HDM parameter points that are in agreement
with the applied constraints.
The type~II parameter points are shown
in blue, and the parameter points of type~IV
are shown in orange. The expected and observed
cross section limits 
\GW{obtained}
by CMS are indicated
\GW{by}
the black dashed and solid lines, respectively,
and the $1\sigma$ and $2\sigma$ uncertainty intervals
are indicated 
\GW{by}
the green and yellow
bands, respectively~\cite{CMSnew}.
The value of
$\mu_{\gamma\gamma}^{\rm exp}$ and its uncertainty
is shown with the magenta error bar at the mass
\GW{value}
at which the excess is most pronounced.
One can see that both types of the S2HDM
considered here can accommodate the observed
excess. As expected from the discussion
in \refse{sec:quanti},
type~II 
\GW{can give rise to larger predicted}
values
of $\mu_{\gamma\gamma}$ due to the additional
suppression of the $h_{95} \to \tau^+ \tau^-$
decay mode. 
\GW{The points featuring the largest values of $\mu_{\gamma\gamma}$ in
type~II are seen to exceed the observed limit of the new CMS analysis 
(which is not applied as a constraint via \texttt{HiggsBounds} in this 
plot). On the other hand, both type~II and type~IV give rise to 
predictions for $\mu_{\gamma\gamma}$ that are very well compatible with 
the new} 
experimental value of $\mu_{\gamma\gamma}^{\rm exp}$
\GW{obtained by CMS}
after the inclusion of the second- and third-year
Run~2 data.\footnote{\GW{As discussed above,} in
type~I and type~III no significant enhancement
of the di-photon branching ratio of
$h_{95}$ is possible, and one finds
$\mu_{\gamma\gamma} \approx \mu_{bb}
\lesssim c_{h_{95} VV}^2$.
Thus, $\mu_{\gamma\gamma}$-values 
\GW{close to}
$\mu_{\gamma\gamma}^{\rm exp}$ require
values of $c_{h_{125}VV}^2 \approx
1 - c_{h_{95}VV}^2$ that are in
significant tension
with the coupling measurements of $h_{125}$.}

\subsection{Combined description of the
excesses}

\begin{figure*}[t]
\centering
\includegraphics[width=0.82\columnwidth]{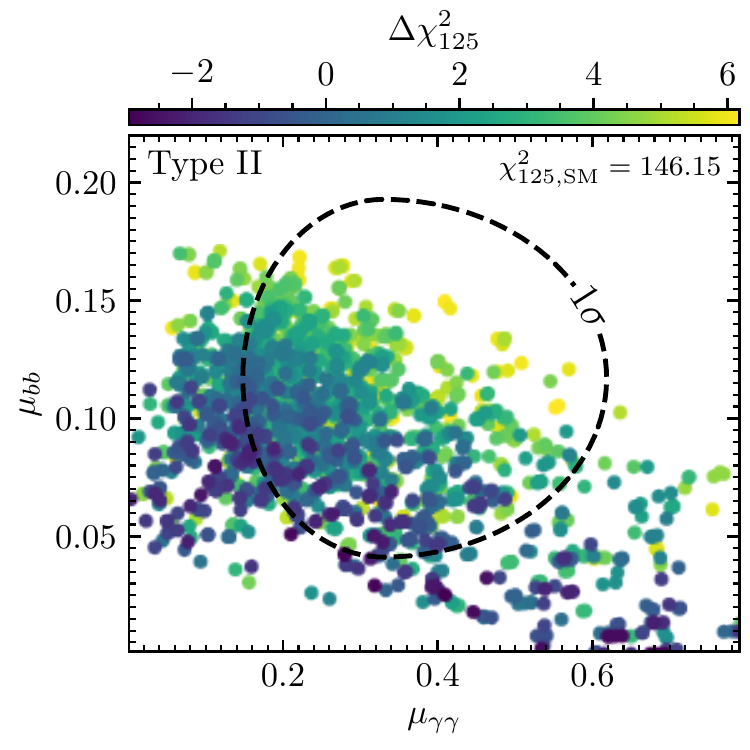}~~~
\includegraphics[width=0.82\columnwidth]{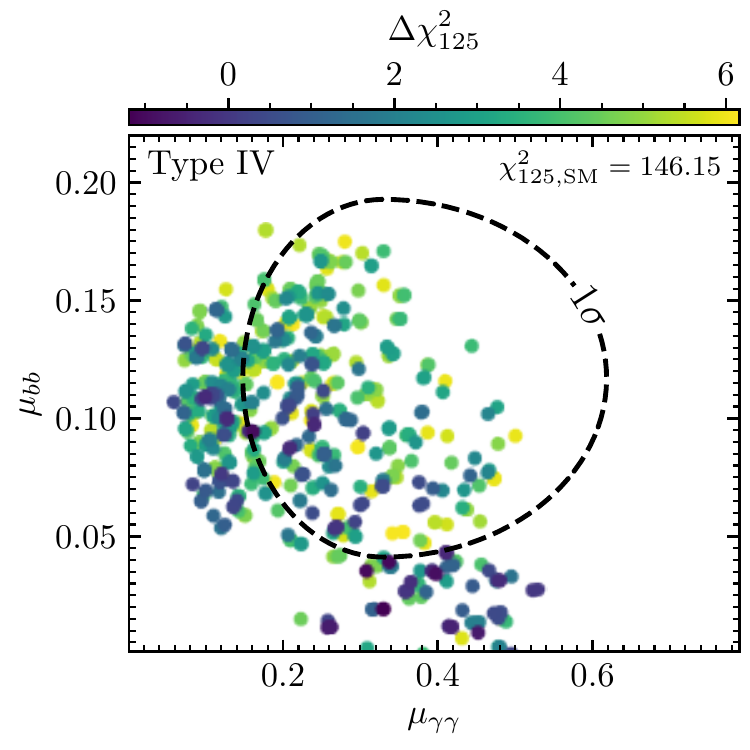}\\[0.4em]
\includegraphics[width=0.82\columnwidth]{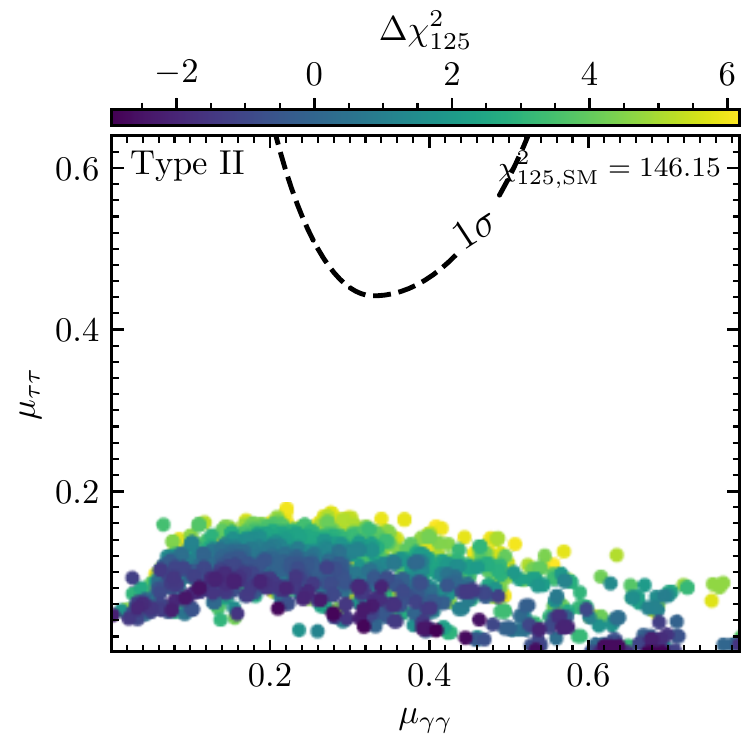}~~~
\includegraphics[width=0.82\columnwidth]{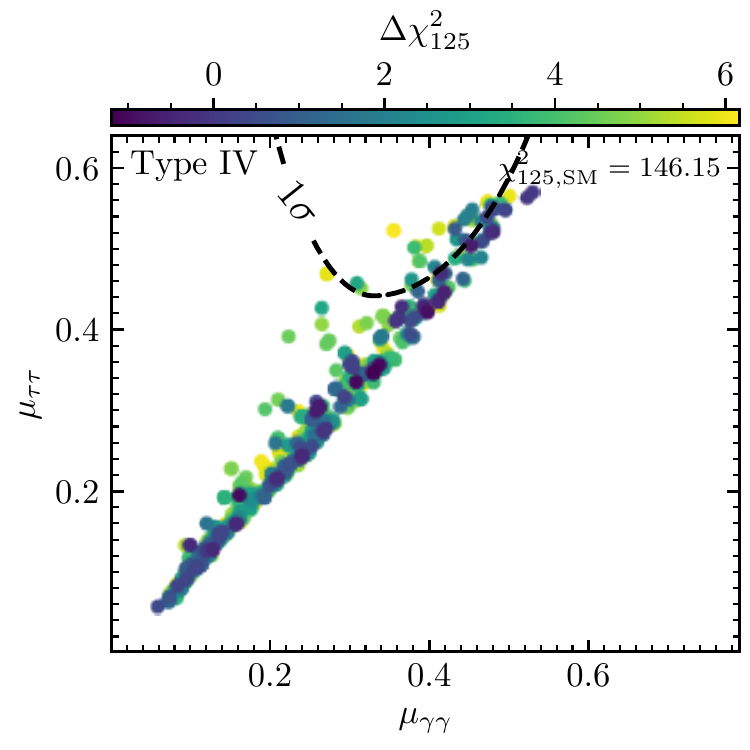}
\vspace*{-0.4cm}
\caption{\small
S2HDM parameter points 
\GW{passing the applied constraints}
in the
\plane{\mu_{\gamma\gamma}}{\mu_{bb}}
(top row) and the
\plane{\mu_{\gamma\gamma}}{\mu_{\tau\tau}}
(bottom row) for type~II (left)
and type~IV (right). The colors of the
points indicate the value of $\Delta \chi^2_{125}$.
The black dashed lines 
indicate the
regions \GW{in} which the two excesses considered
in each plot are accommodated at a level of
$1\sigma$ or better, i.e.~$\chi^2_{\gamma\gamma}
+ \chi^2_{bb} \leq 2.3$ (top row) and
$\chi^2_{\gamma\gamma}
+ \chi^2_{\tau\tau} \leq 2.3$ (bottom row).}
\label{fig:yyllbb}
\end{figure*}

We demonstrated in the previous section that
both \GW{the Yukawa types II and IV}
can describe the
excess in the di-photon channel observed
by CMS. Now we turn to the question whether
additionally also the $b \bar b$ excess
observed at LEP and the $\tau^+ \tau^-$
excess at CMS can be accommodated.

Starting with the $b \bar b$ excess,
we show in the top row of \reffi{fig:yyllbb}
the parameter points 
\GW{passing the applied constraints}
in the 
\plane{\mu_{\gamma\gamma}}{\mu_{b b}}.
The parameter points of type~II and type~IV
are shown in left and the right plot, respectively.
The colors of the points indicate the
value of $\Delta \chi^2_{125}$ showing
the compatibility with the LHC rate
measurements of $h_{125}$.
The black dashed lines 
indicate the
region in which the excesses are described
at a level of $1\sigma$ or better,
i.e.~$\chi^2_{\gamma\gamma} +
\chi^2_{bb} \leq 2.3$ 
(see \refeq{eq:chisq95indi}).
The shape of these lines is asymmetrical
due to the asymmetrical
uncertainties of $\mu_{\gamma\gamma}^{\rm exp}$
used in the definition of $\chi^2_{\gamma\gamma}$
in \refeq{eq:chisq95indi}.

One can see that we find points
inside the $1\sigma$ preferred region 
in the upper left and 
right plots.
Thus, both type~II and type~IV
are able to describe the di-photon excess
and the $b \bar b$ excess
simultaneously.
At the same time the properties
of the second-lightest scalar
$h_{125}$ are such that the
LHC rate measurements
can be accommodated 
\GW{at the same $\chi^2$ level}
as
in the SM,
i.e.~$\Delta \chi^2_{125} \approx 0$,
or even marginally
better, i.e.~$\Delta \chi^2_{125} < 0$.
At the current level of experimental
precision, the description of both
excesses is therefore possible in combination
with the presence of
a Higgs boson at $125\gev$ that
would so far be indistinguishable from
a SM Higgs boson.

Turning to the di-tau excess,
we show in the bottom row of
\reffi{fig:yyllbb} the 
parameter points 
\GW{passing the applied constraints}
in the \plane{\mu_{\gamma\gamma}}{\mu_{\tau\tau}}.
As before, the colors of the points
indicate the values of $\Delta \chi^2_{125}$,
and the black dashed lines 
indicate the region in which the di-photon excess and
the di-tau excess are described at a level
of $1\sigma$ or better, i.e.~$\chi^2_{\gamma
\gamma} + \chi^2_{\tau\tau} \leq 2.3$.

In the lower left plot, showing
the parameter points of the scan in
type~II, one can see that there are no
points within or close to the black line. 
This finding is in agreement with the
\GW{discussion} in \refse{sec:quanti}.
It is also \SH{qualitatively} unchanged as compared to
the results of \citere{Biekotter:2022jyr},
where $\mu_{\gamma\gamma}^{\rm exp} = 0.6 \pm 0.2$
was used: the new and somewhat lower experimental 
\GW{central}
value of $\mu_{\gamma\gamma}^{\rm exp}$ has no impact on the
(non-)compatibility of the $\gamma\gamma$ and the
$\tau^+\tau^-$ excesses in Yukawa type~II.

The lower right plot shows
the 
parameter points 
\GW{passing the applied constraints}
from the
scan in type~IV. One can observe that
the values of $\mu_{\tau\tau}$ overall increase
with increasing value of $\mu_{\gamma\gamma}$.
The parameter points that predict
the largest values for the signal rates
reach the lower edge of the black line 
that indicates the preferred region
regarding the two excesses.
However, even these points lie substantially
below the central value of $\mu_{\tau\tau}^{\rm exp}$.
A simultaneous description
of both excesses at $95\gev$ observed by CMS
is therefore possible only at the
level of $1\,\sigma$ at best.
Although larger values of $\mu_{\tau\tau}$
are theoretically possible in
type~IV~\cite{Biekotter:2022jyr},
the application of cross-section limits
from Higgs-boson searches exclude such
parameter points.
These constraints arise in particular 
from recent searches performed by CMS 
for the production of a Higgs boson
in association with a top-quark pair or
in association with a $Z$~boson, with subsequent
decay into tau pairs~\cite{CMS-PAS-EXO-21-018}.

Constraints on the interpretation
of the di-tau excess 
\GW{as an additional Higgs boson}
were also derived from
cross-section measurements of the Higgs boson
at $125\gev$.
In particular, \citere{Iguro:2022dok}
investigated the sensitivity of the ATLAS
measurement assuming the production
of $h_{125}$ in association
with a top-quark pair and subsequent decay
into di-tau
pairs~\cite{ATLAS:2022yrq}.\footnote{In
\citere{Coloretti:2023wng} the invariant-mass
spectra of the $h_{125} \to WW^*$ decay
channel measured by ATLAS~\cite{ATLAS:2022ooq}
and CMS~\cite{CMS:2022uhn}
were considered. However, the decay of
$h_{95} \to WW^*$ is highly off-shell,
suppressing the corresponding branching
ratio by orders of magnitude compared to
the one of $h_{125}$.
As a result, there is no sensitivity
in this decay channel to the presence
of $h_{95}$ according to our model
interpretation of the excesses.}
The ATLAS analysis considered
an invariant di-tau mass in the range between
$50\gev$ and $200\gev$ and is based
on the full Run~2 data \GW{set}. 
We emphasize, however, that the
constraints extracted
in \citere{Iguro:2022dok}
are affected by the lack of publicly available
information on
the correlations between the different mass bins.

In summary, the S2HDM type~II can
simultaneously describe
the CMS di-photon excess and the $b \bar b$
excess observed at LEP, whereas no significant
\GW{contribution to the
signal strength of} the CMS di-tau excess is 
\GW{generated.}
In type~IV, in addition also a 
\GW{contribution to the
di-tau signal strength can occur}, although
the largest possible signal rates of about
$\GW{\mu_{\tau\tau}} = 0.5$ are somewhat
below the 
\GW{experimentally preferred}
range of $\mu_{\tau\tau}^{\rm exp}
= 1.2 \pm 0.5$.

\medskip
\SH{Our results in the S2HDM}
\GW{can be generalised to other 
extended Higgs sectors containing at least a
second Higgs doublet and
at least one scalar singlet.
Our analysis indicates that
the conclusions in various models
that have previously been considered
as an explanation for the di-photon excess are expected to be affected by the
modified value
of $\mu_{\gamma\gamma}^{\rm exp}$. 
This applies in particular to}
\SH{supersymmetric extensions
of the SM, which were shown to be able to accommodate
a signal at about $95\gev$ with
a signal strength that in most cases
was predicted to be
at the lower end of
the previous
$\mu_{\gamma\gamma}^{\rm exp}$-range~\cite{Biekotter:2017xmf,
Domingo:2018uim,
Hollik:2018yek,
Biekotter:2019gtq,
Choi:2019yrv,
Cao:2019ofo,
Biekotter:2021qbc}.
Requiring also agreement with the LEP excess
resulted in $\mu_{\gamma\gamma} \approx
0.3$~\cite{Domingo:2018uim,Biekotter:2019gtq,Biekotter:2021qbc},
which}
\GW{turns out to 
be in very good
agreement with the updated result from CMS.}

\subsection{Prospects at future colliders}

We finally discuss how future collider
experiments will shed light on the
possible presence
of a Higgs boson below $125\gev$ as considered here.
In the S2HDM the mixing between the singlet-like
state at $95\gev$ and the SM-like state
at $125\gev$ determines the strengths of the
couplings of $h_{95}$ to fermions and
gauge bosons. Thus,
in addition to directly searching for
$h_{95}$, a
complementary -- although more model-dependent -- strategy consists in
the search for modifications of the cross sections
of $h_{125}$ compared to the ones of
a SM Higgs boson. We start with discussing this
approach in the following.

Currently, the experimental precision
of the observed couplings of $h_{125}$
\GW{is} at the level of ten to twenty
percent~\cite{CMS:2022dwd,ATLAS:2022vkf}.
During the high-luminosity phase of the LHC \GW{(HL-LHC)},
the experimental precision of these couplings
\GW{is expected to be reduced}
to the level of
a few percent~\cite{Cepeda:2019klc}.\footnote{Here
it is assumed that no undetected decay mode
of $h_{125}$ 
\GW{into} BSM particles is
present.}
A future $e^+ e^-$ collider with sufficient
energy to produce $h_{125}$
could further improve
the experimental precision to the
sub-percent level.
As an example, we will consider
in the following the expected precision
of the International Linear Collider
(ILC) operating at a center-of-mass
energy of $250\gev$ and collecting
$2~\mathrm{ab}^{-1}$ of
integrated luminosity~\cite{Bambade:2019fyw}.
We note that here and in the following 
the specific example of the projections for the 
ILC is meant to showcase
the potential impact of the coupling
measurements at a future $e^+ e^-$ collider.
In fact, very similar results would be obtained
considering the other proposals for a
``Higgs factory'' operating at about~250~GeV, such
as CLIC, CEPC or the FCC-ee~\cite{deBlas:2019rxi}.

\begin{figure}[t]
\centering
\includegraphics[width=0.96\columnwidth]{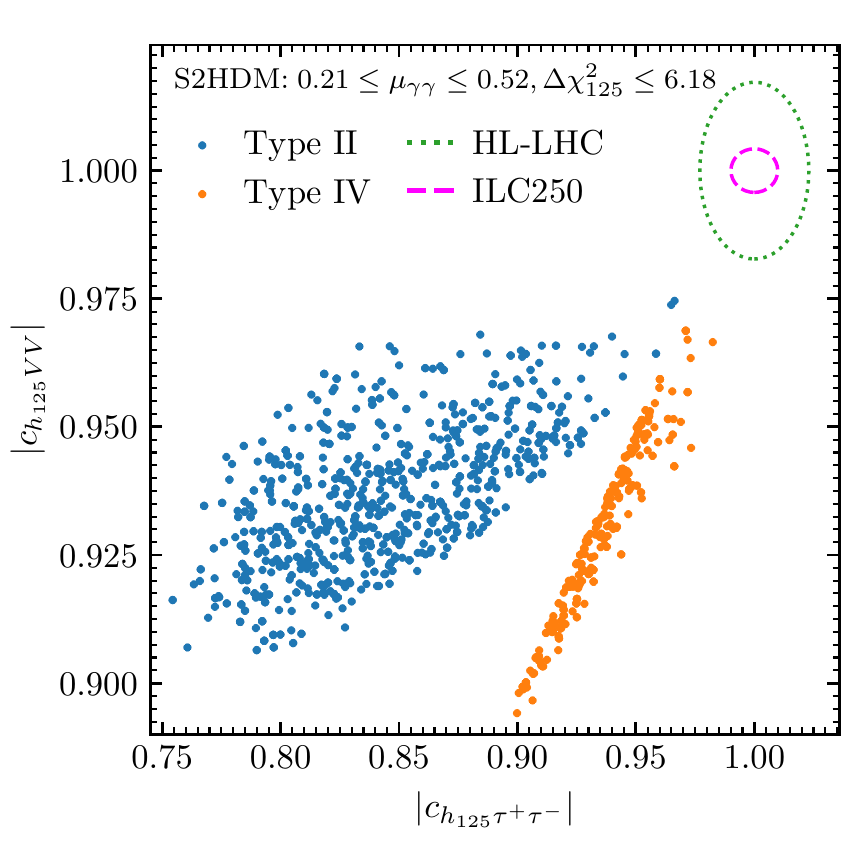}
\vspace*{-0.4cm}
\caption{\small
S2HDM parameter points 
\GW{passing the applied constraints}
that
predict a di-photon signal strength
in the 
\GW{preferred range of $0.21 \leq \mu_{\gamma\gamma}
\leq 0.52$
in view of the excess observed by}
CMS~\cite{CMSnew} in the
\plane{|c_{h_{125} \tau^+ \tau^-}|}{|c_{h_{125} VV}|}.
The type~II and the type~IV parameter
points are shown in blue and orange,
respectively.
The green dotted and the magenta
dashed ellipses indicate the 
projected experimental precision of the
coupling measurements at the
HL-LHC~\cite{Cepeda:2019klc} and the
ILC250~\cite{Bambade:2019fyw}, respectively,
with their centers located at the SM values.}
\label{fig:h125cpls}
\end{figure}

In \reffi{fig:h125cpls} we show
the 
parameter points 
\GW{passing the applied constraints}
of the scan in type~II (blue)
and in type~IV (orange) that
provide a good description
of the di-photon excess,
i.e.~$0.21 \leq \mu_{\gamma\gamma} \leq 0.52$,
in the \plane{|c_{h_{125} \tau^+ \tau^-}|}{|c_{h_{125} VV}|}.
Here $c_{h_{125} \tau^+ \tau^-}$
and $c_{h_{125} VV}$
are the effective coefficients of the
coupling of $h_{125}$ to tau-leptons
and the gauge bosons $V=Z,W$, respectively.
These coefficients are normalized such that they
are equal to one in the SM.
Centered at the SM prediction, we also
indicate with the green dotted ellipse the
expected precision on the coupling coefficients
after the 
\GW{HL-LHC}
will have collected $3000~\mathrm{fb}^{-1}$
of integrated luminosity.
Finally, the magenta dashed ellipse indicates
the expected experimental precision
after a combination of the HL-LHC data
and the ILC data collected 
at $\sqrt{s} = 250 \gev$ (ILC250) with an 
integrated luminosity of $2~\mathrm{ab}^{-1}$.
We note that these experimental
projections have been obtained assuming
that the cross section measurements are
according to the predictions of the SM.

One can see that the points of both
types all lie outside of the green ellipse.
For the points with the largest deviations
from the SM, the anticipated HL-LHC precision
would be sufficient to distinguish between 
\GW{SM-like properties of $h_{125}$ and the 
predictions of the S2HDM for parameter regions that are in accordance 
with the observed di-photon excess}.
However, for the \GW{S2HDM points that are} closest to the 
SM value, no distinction at the $2\,\sigma$ 
level could be established. Consequently, the
HL-LHC will not be able to entirely probe
the S2HDM interpretation of the di-photon
excess at $95\gev$ \GW{based on the coupling measurements of $h_{125}$}.
Moreover, 
\GW{for many of the displayed blue and orange points the 
expected HL-LHC precision, indicated by the}
size of the green ellipse,
will
not be sufficient to distinguish between
a type~II and a type~IV interpretation.

Now we compare the model predictions with
the expected precision at the ILC250,
indicated by the magenta ellipse.
One can see that under the assumption that
no modifications of the properties of
$h_{125}$ will be observed even at the
ILC, all parameter points would be excluded
with 
\GW{high} experimental significance.
\GW{On the other hand,}
for each point in the S2HDM 
describing the di-photon excess, 
\GW{a clear deviation of the properties of $h_{125}$ from the SM 
predictions could be established via the}
coupling measurements.
The ILC also has a significantly larger
potential to distinguish between a type~II
and a type~IV scenario, although even the
ILC precision might not be sufficient to
distinguish between the types for the
parameter points with the largest values
of $c_{h_{125} \tau^+ \tau^-}$
and $c_{h_{125} VV}$.
\GW{Information about} the direct production of
$h_{95}$ and its coupling measurements
\GW{will of course be instrumental to further}
probe the S2HDM scenarios. 

In our S2HDM interpretation of the di-photon
excess, $h_{95}$ is required to have a
non-vanishing coupling to
top quarks, and thus also to gauge bosons,
in order to be the origin of this excess.
Moreover, a sizable coupling of $h_{95}$
\GW{to the $Z$ boson}
is required if this state is also supposed to
be the origin of the $b \bar b$ excess
observed at LEP.
In this case, a future lepton collider running at
$250\gev$ has the capability to produce
$h_{95}$ in large numbers~\cite{Drechsel:2018mgd,
Wang:2020lkq}.
From the resulting
cross-section measurements, the couplings
of $h_{95}$ 
\GW{could}
be determined with a
precision that is expected to greatly improve on
the precision achievable at
the~LHC.\footnote{Experimental
projections for Higgs coupling measurements at the HL-LHC
are only publicly available for the discovered Higgs boson 
at 125~GeV. In contrast to the cleaner experimental environment at 
an $e^+ e^-$ collider, at the LHC it is not feasible
to obtain projections for the accuracy of coupling measurements 
for additional Higgs bosons without detailed simulations
taking into account systematical uncertainties. Since such a dedicated
simulation would be beyond the scope of the present paper, we do not attempt 
to provide precise quantitative estimates for the achievable accuracy 
on the couplings of $h_{95}$ at the HL-LHC.
However, a rough estimate of the precision
for the signal rates in the di-photon and
di-tau channel assuming $3~\mathrm{ab}^{-1}$
can be achieved by a simple rescaling
with the square root of the luminosity,
yielding a precision of about 
10\% for the di-photon and the di-tau channel.}
Thus, if a new state at $95\gev$ exists,
a future $e^+ e^-$ collider such as the
ILC 
\GW{is expected to}
be of vital importance for the
determination of the underlying model
that is realized in nature.

\begin{figure}[t]
\centering
\includegraphics[width=0.96\columnwidth]{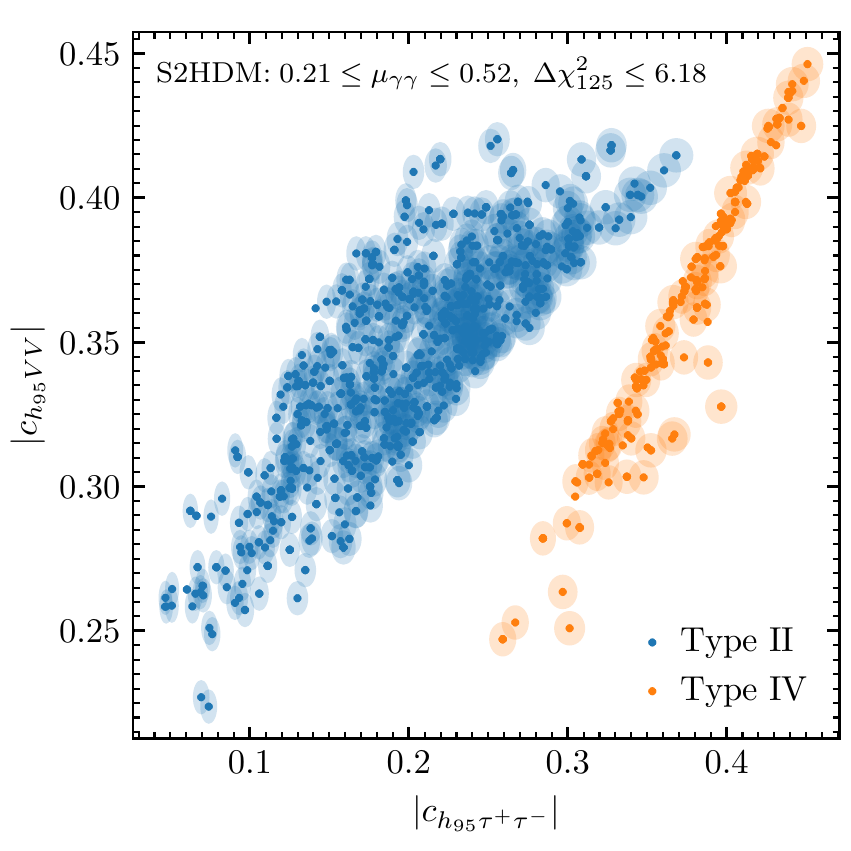}
\vspace*{-0.4cm}
\caption{\small
S2HDM parameter points passing the
applied constraints that
predict a di-photon signal strength
in the preferred range $0.21 \leq \mu_{\gamma\gamma}
\leq 0.52$
in view of the excess observed by
CMS~\cite{CMSnew} in the
\plane{|c_{h_{95} \tau^+ \tau^-}|}{|c_{h_{95} VV}|}.
The type~II and the type~IV parameter
points are shown in blue and orange,
respectively. The shaded ellipses around
the dots indicate the projected experimental
precision with which the couplings of
$h_{95}$ could be measured at the ILC250 
\GW{with}
$2~\mathrm{ab}^{-1}$ of integrated
luminosity, which we evaluated according to 
\citere{Heinemeyer:2021msz}.}
\label{fig:h95cpls}
\end{figure}

In order to showcase the potential
of the ILC for
discriminating different models that
give rise to the state at $h_{95}$,
we show in \reffi{fig:h95cpls}
the parameter points
of our scans in the
\plane{|c_{h_{95} \tau^+ \tau^-}|}{|c_{h_{95} VV}|}.
Here, $c_{h_{95} \tau^+ \tau^-}$
and $c_{h_{95} VV}$ are the effective
coefficients for the couplings of $h_{95}$
to tau-leptons and gauge bosons, respectively.
These coefficients are normalized
such that they are equal to one for a
hypothetical SM Higgs boson at the mass
of $h_{95}$. As in \reffi{fig:h125cpls},
the parameter points of
type~II and type~IV are shown in
blue and orange, respectively,
and we only depict the parameter points
that provide a good description of
the di-photon excess observed by CMS.
In addition to the theoretical prediction
of the coupling coefficients, indicated
with the dots, we also indicated the
experimental precision with which the
respective couplings could be measured
at the ILC 
\GW{by means of} the shaded ellipses
around each dot.
We estimated the experimental precision
of the coupling measurements 
\GW{for the ILC250 with $2~\mathrm{ab}^{-1}$
of integrated luminosity}
according
to the approach discussed in
\citere{Heinemeyer:2021msz}.

One can observe in \reffi{fig:h95cpls}
that the blue points
and the orange points are clearly
separated from each other.
For a fixed value of the gauge-boson
coupling, the parameter
points of type~IV predict larger couplings
to tau-leptons compared to the parameter
points of type~II. This is in line with
the discussion in \refse{sec:quanti}: In type~II
one has
$c_{h_{95} \tau^+ \tau^-} = c_{h_{95} b \bar b}$,
such that the enhancement of the di-photon
branching ratio via the condition
$|c_{h_{95} t \bar t} / c_{h_{95} b \bar b}| > 1$
is achieved in the regime
in which $c_{h_{95} \tau^+ \tau^-}$
is suppressed. On the other hand,
in type~IV one has
$c_{h_{95} \tau^+ \tau^-} = c_{h_{95} t \bar t}$,
such that the coupling to tau-leptons
is less suppressed in the regime in which the
di-photon branching ratio is enhanced.

As a consequence of the separation of
the points of the two types, combined with the 
high anticipated precision of the $h_{95}$ coupling 
measurements at the ILC250, there are
no blue or orange ellipses that overlap.
Thus, the coupling
measurements of $h_{95}$ at the ILC
would be sufficient to distinguish
between a type~II or a type~IV interpretation.
In combination with the experimental
observation regarding $h_{125}$ (see
discussion above), a lepton collider like
the ILC would be able to 
\GW{scrutinize} the underlying physics model that is realized
in nature.

\section{Conclusions and outlook}
\label{sec:conclu}
Recently, upon the inclusion of the
full Run~2 data set and substantially
refined analysis techniques,
the CMS collaboration has
confirmed an excess of \GW{about} $3\,\sigma$
local significance at about $95\gev$
in the low-mass Higgs boson searches
in the di-photon final state.
\GW{An excess at this mass value with similar significance}
had previously been reported
based on the $8\tev$ Run~1 and the first-year
Run~2 data set.
We have 
\GW{investigated the interpretation of this excess as a 
di-photon resonance arising from the production of} 
a Higgs boson in the
\GW{Two-Higgs doublet model that is extended by a}
complex singlet 
(S2HDM).
We have shown that a good description of the
excess is possible in the Yukawa
type~II and~IV, while being in agreement
with all other collider searches for additional
Higgs bosons, the measurements of the 
\GW{properties of the}
SM-like
Higgs boson at $125\gev$,
\GW{and further experimental and theoretical constraints}. 
At the same time,
the model can account for all or a large
fraction of the observed 
DM relic abundance in agreement with the measurements
of the Planck satellite.

Previously, a signal strength
for the di-photon excess observed by CMS 
of $\mu_{\gamma\gamma}^{\rm exp} = 0.6 \pm 0.2$
had been obtained 
utilizing the 
\GW{data from the
first year of Run~2 and of Run~1}.
This \GW{relatively high central value of the signal strength 
gave rise to a} 
preference to
a type~II Yukawa structure,
in which larger signal rates
of the state at $95\gev$ can be achieved
compared to the type~IV.
After the inclusion of the remaining
Run~2 data
and 
\GW{performing various improvements of the}
experimental analysis, 
the new CMS 
\GW{result 
shows an excess with a local significance that is essentially unchanged 
compared to the previous result but which yields an
interpretation in terms of a smaller central value of the signal 
strength with reduced uncertainties,
$\mu_{\gamma\gamma}^{\rm exp} =
\muCMS^{+\dmuCMSpl}_{-\dmuCMSmi}$}.
We have shown that
as a result of the smaller \GW{central} value of
$\mu_{\gamma\gamma}^{\rm exp}$ 
\GW{both Yukawa types provide an equally well description} 
of the di-photon excess in the S2HDM.

The di-photon excess observed at CMS
is especially intriguing in view of
additional excesses that appeared at
approximately the same mass.
An excess of events above the SM
expectation with \GW{about} $2\,\sigma$ local
significance was observed at LEP in searches
\GW{for}
Higgsstrahlung production
of a scalar state that then decays to a pair
of bottom quarks. Moreover, CMS observed
an excess with \GW{about} $3\,\sigma$ local significance
consistent with a mass of about
$95\gev$ in searches 
\GW{for}
the production
of a Higgs boson via gluon fusion and subsequent
decay into \GW{tau} pairs.

We have demonstrated that the S2HDM type~II can
simultaneously describe
the CMS di-photon excess and the $b \bar b$
excess observed at LEP, whereas no significant
signal for the CMS di-tau excess is possible 
\GW{in this model}.
In the S2HDM type~IV, on the other hand, in addition
also a sizable signal 
\GW{strength in the di-tau channel can occur}.
However,
even in type~IV
the 
\GW{maximally reachable}
signal rates are
smaller than the
signal strengths
that \GW{would be}
required to describe the
di-tau excess at the level of~$1\,\sigma$.

Our analysis in the S2HDM serves as
an example study from which more 
\GW{general}
conclusions valid for a wider class of
extensions of the SM can be drawn.
Notably, supersymmetric extensions
were previously shown to be able to accommodate a
di-photon signal at about $95\gev$ that 
\GW{turns out to}
be in good agreement with the updated
experimental value of $\mu_{\gamma\gamma}^{\rm exp}$.

In the near future,
\GW{the possible presence}
of a Higgs boson at $95\gev$
\GW{can be directly tested by the eagerly awaited results from the}
corresponding \mbox{ATLAS} searches in the di-photon
and the di-tau final states covering the
mass region below $125\gev$ and utilizing
the full Run~2 data.
\GW{Further into the future,}
the scenarios 
\GW{discussed} here will
be tested in a twofold way at future Runs
of the (HL)-LHC, where the direct searches
for $h_{95}$ and the coupling measurements
of $h_{125}$ will benefit 
\GW{in particular from a significant}
increase of statistics.
Nevertheless, we have shown that the experimental
precision of the coupling measurements of
the Higgs boson at $125\gev$ might not be
sufficient to exclude the S2HDM interpretation
of the excesses at $95\gev$, or conversely 
confirm a deviation from the SM predictions.

Going beyond the (HL-)LHC projections,
we have discussed the experimental prospects
at a future $e^+ e^-$ collider, considering
as an example the ILC operating
at $250\gev$ with an integrated luminosity 
of $2~\mathrm{ab}^{-1}$. At the ILC250,
the couplings of $h_{125}$ could be determined
in an effectively model independent way at
sub-percent level precision.
Assuming that no deviations from the SM
predictions would be observed, the measurements
of the couplings of $h_{125}$ would
significantly disfavour
the S2HDM interpretation of the excess
at $95\gev$.
Conversely, 
\GW{a clear deviation from the SM predictions will be established}
if the coupling measurements of $h_{125}$
will be according to the predictions
of any S2HDM parameter
point describing the excess.

Although the possible state
at $95\gev$ has suppressed couplings compared
to $h_{125}$, the ILC could produce
$h_{95}$ in large numbers 
\GW{if it has a sufficiently large coupling to $Z$ bosons}.
We have shown that
the clean environment of an $e^+e^-$ collider
would allow for a determination of
the couplings of $h_{95}$
at percent-level precision.
As such, we demonstrated that the ILC,
in contrast to the HL-LHC,
could distinguish between a type~II and
a type~IV description of the excesses.

\section*{Acknowledgements}
G.W.~acknowledges support by the Deutsche
Forschungsgemeinschaft (DFG, German Research
Foundation) under Germany‘s Excellence
Strategy -- EXC 2121 ``Quantum Universe'' --
390833306.
The work of G.W.~has been partially funded
by the Deutsche Forschungsgemeinschaft 
(DFG, German Research Foundation) - 491245950. 
S.H~acknowledges support from the grant IFT
Centro de Excelencia Severo Ochoa
CEX2020-001007-S funded by 
MCIN/AEI/10.13039/501100011033.
The work of S.H.~was supported in part by the
grant PID2019-110058GB-C21 funded by
MCIN/AEI/10.13039/501100011033 and by
``ERDF A way of making Europe''.
The work of T.B.~is supported by the German
Bundesministerium f\"ur Bildung und Forschung
(BMBF, Federal Ministry of Education and Research)
-- project 05H21VKCCA.

\bibliographystyle{JHEP}
\bibliography{refs}

\providecommand{\href}[2]{#2}\begingroup\raggedright\begin{thebibliography}{10}

\bibitem{Aad:2012tfa}
{\scshape ATLAS} collaboration, \emph{{Observation of a new particle in the
  search for the Standard Model Higgs boson with the ATLAS detector at the
  LHC}}, \href{https://doi.org/10.1016/j.physletb.2012.08.020}{\emph{Phys.
  Lett. B} {\bfseries 716} (2012) 1}
  [\href{https://arxiv.org/abs/1207.7214}{{\ttfamily 1207.7214}}].

\bibitem{Chatrchyan:2012xdj}
{\scshape CMS} collaboration, \emph{{Observation of a New Boson at a Mass of
  125 GeV with the CMS Experiment at the LHC}},
  \href{https://doi.org/10.1016/j.physletb.2012.08.021}{\emph{Phys. Lett. B}
  {\bfseries 716} (2012) 30} [\href{https://arxiv.org/abs/1207.7235}{{\ttfamily
  1207.7235}}].

\bibitem{CMS:2022dwd}
{\scshape CMS} collaboration, \emph{{A portrait of the Higgs boson by the CMS
  experiment ten years after the discovery}},
  \href{https://doi.org/10.1038/s41586-022-04892-x}{\emph{Nature} {\bfseries
  607} (2022) 60} [\href{https://arxiv.org/abs/2207.00043}{{\ttfamily
  2207.00043}}].

\bibitem{ATLAS:2022vkf}
{\scshape ATLAS} collaboration, \emph{{A detailed map of Higgs boson
  interactions by the ATLAS experiment ten years after the discovery}},
  \href{https://doi.org/10.1038/s41586-022-04893-w}{\emph{Nature} {\bfseries
  607} (2022) 52} [\href{https://arxiv.org/abs/2207.00092}{{\ttfamily
  2207.00092}}].

\bibitem{Abbiendi:2002qp}
{\scshape OPAL} collaboration, \emph{{Decay mode independent searches for new
  scalar bosons with the OPAL detector at LEP}},
  \href{https://doi.org/10.1140/epjc/s2002-01115-1}{\emph{Eur. Phys. J. C}
  {\bfseries 27} (2003) 311}
  [\href{https://arxiv.org/abs/hep-ex/0206022}{{\ttfamily hep-ex/0206022}}].

\bibitem{Barate:2003sz}
{\scshape LEP Working Group for Higgs boson searches, ALEPH, DELPHI, L3, OPAL}
  collaboration, \emph{{Search for the standard model Higgs boson at LEP}},
  \href{https://doi.org/10.1016/S0370-2693(03)00614-2}{\emph{Phys. Lett. B}
  {\bfseries 565} (2003) 61}
  [\href{https://arxiv.org/abs/hep-ex/0306033}{{\ttfamily hep-ex/0306033}}].

\bibitem{Schael:2006cr}
{\scshape ALEPH, DELPHI, L3, OPAL, LEP Working Group for Higgs Boson Searches}
  collaboration, \emph{{Search for neutral MSSM Higgs bosons at LEP}},
  \href{https://doi.org/10.1140/epjc/s2006-02569-7}{\emph{Eur. Phys. J. C}
  {\bfseries 47} (2006) 547}
  [\href{https://arxiv.org/abs/hep-ex/0602042}{{\ttfamily hep-ex/0602042}}].

\bibitem{Group:2012zca}
{\scshape CDF, D0} collaboration, \emph{{Updated Combination of CDF and D0
  Searches for Standard Model Higgs Boson Production with up to 10.0 fb$^{-1}$
  of Data}},  7, 2012 [\href{https://arxiv.org/abs/1207.0449}{{\ttfamily
  1207.0449}}].

\bibitem{CMS:2015ocq}
{\scshape CMS} collaboration, \emph{{Search for new resonances in the diphoton
  final state in the mass range between 80 and 110 GeV in pp collisions at
  $\sqrt{s}=8$ TeV}},  Tech. Rep.
  \href{https://cds.cern.ch/record/2063739}{CMS-PAS-HIG-14-037} (2015).

\bibitem{CMS:2018cyk}
{\scshape CMS} collaboration, \emph{{Search for a standard model-like Higgs
  boson in the mass range between 70 and 110 GeV in the diphoton final state in
  proton-proton collisions at $\sqrt{s}=$ 8 and 13 TeV}},
  \href{https://doi.org/10.1016/j.physletb.2019.03.064}{\emph{Phys. Lett. B}
  {\bfseries 793} (2019) 320}
  [\href{https://arxiv.org/abs/1811.08459}{{\ttfamily 1811.08459}}].

\bibitem{CMS:2018rmh}
{\scshape CMS} collaboration, \emph{{Search for additional neutral MSSM Higgs
  bosons in the $\tau\tau$ final state in proton-proton collisions at
  $\sqrt{s}=$ 13 TeV}},
  \href{https://doi.org/10.1007/JHEP09(2018)007}{\emph{JHEP} {\bfseries 09}
  (2018) 007} [\href{https://arxiv.org/abs/1803.06553}{{\ttfamily
  1803.06553}}].

\bibitem{ATLAS:2018xad}
{\scshape ATLAS} collaboration, \emph{{Search for resonances in the 65 to 110
  GeV diphoton invariant mass range using 80 fb$^{-1}$ of $pp$ collisions
  collected at $\sqrt{s}=13$ TeV with the ATLAS detector}},  Tech. Rep.
  \href{https://cds.cern.ch/record/2628760}{ATLAS-CONF-2018-025} (7, 2018).

\bibitem{CMS:2022goy}
{\scshape CMS} collaboration, \emph{{Searches for additional Higgs bosons and
  for vector leptoquarks in $\tau\tau$ final states in proton-proton collisions
  at $\sqrt{s}$ = 13 TeV}},  \href{https://arxiv.org/abs/2208.02717}{{\ttfamily
  2208.02717}}.

\bibitem{ATLAS:2022abz}
{\scshape ATLAS} collaboration, \emph{{Search for boosted diphoton resonances
  in the 10 to 70 GeV mass range using 138 fb$^{-1}$ of 13 TeV $pp$ collisions
  with the ATLAS detector}},
  \href{https://arxiv.org/abs/2211.04172}{{\ttfamily 2211.04172}}.

\bibitem{CMSnew}
{\scshape CMS} collaboration, \emph{{Search for low mass resonances in the
  diphoton final state in proton-proton collisions at $\sqrt{s}=$13 TeV with
  the full Run~2 dataset}},  Tech. Rep.
  \href{https://cds.cern.ch/record/2852907}{CMS-HIG-20-002} (2023).

\bibitem{Cao:2016uwt}
J.~Cao, X.~Guo, Y.~He, P.~Wu and Y.~Zhang, \emph{{Diphoton signal of the light
  Higgs boson in natural NMSSM}},
  \href{https://doi.org/10.1103/PhysRevD.95.116001}{\emph{Phys. Rev. D}
  {\bfseries 95} (2017) 116001}
  [\href{https://arxiv.org/abs/1612.08522}{{\ttfamily 1612.08522}}].

\bibitem{Fox:2017uwr}
P.J.~Fox and N.~Weiner, \emph{{Light Signals from a Lighter Higgs}},
  \href{https://doi.org/10.1007/JHEP08(2018)025}{\emph{JHEP} {\bfseries 08}
  (2018) 025} [\href{https://arxiv.org/abs/1710.07649}{{\ttfamily
  1710.07649}}].

\bibitem{Richard:2017kot}
F.~Richard, \emph{{Search for a light radion at HL-LHC and ILC250}},
  \href{https://arxiv.org/abs/1712.06410}{{\ttfamily 1712.06410}}.

\bibitem{Haisch:2017gql}
U.~Haisch and A.~Malinauskas, \emph{{Let there be light from a second light
  Higgs doublet}}, \href{https://doi.org/10.1007/JHEP03(2018)135}{\emph{JHEP}
  {\bfseries 03} (2018) 135}
  [\href{https://arxiv.org/abs/1712.06599}{{\ttfamily 1712.06599}}].

\bibitem{Biekotter:2017xmf}
T.~Biek\"otter, S.~Heinemeyer and C.~Mu\~noz, \emph{{Precise prediction for the
  Higgs-boson masses in the $\mu \nu $ SSM}},
  \href{https://doi.org/10.1140/epjc/s10052-018-5978-7}{\emph{Eur. Phys. J. C}
  {\bfseries 78} (2018) 504}
  [\href{https://arxiv.org/abs/1712.07475}{{\ttfamily 1712.07475}}].

\bibitem{Liu:2018xsw}
D.~Liu, J.~Liu, C.E.M.~Wagner and X.-P.~Wang, \emph{{A Light Higgs at the LHC
  and the B-Anomalies}},
  \href{https://doi.org/10.1007/JHEP06(2018)150}{\emph{JHEP} {\bfseries 06}
  (2018) 150} [\href{https://arxiv.org/abs/1805.01476}{{\ttfamily
  1805.01476}}].

\bibitem{Domingo:2018uim}
F.~Domingo, S.~Heinemeyer, S.~Pa\ss{}ehr and G.~Weiglein, \emph{{Decays of the
  neutral Higgs bosons into SM fermions and gauge bosons in the
  $\mathcal{CP}$-violating NMSSM}},
  \href{https://doi.org/10.1140/epjc/s10052-018-6400-1}{\emph{Eur. Phys. J. C}
  {\bfseries 78} (2018) 942}
  [\href{https://arxiv.org/abs/1807.06322}{{\ttfamily 1807.06322}}].

\bibitem{Biekotter:2019kde}
T.~Biek\"otter, M.~Chakraborti and S.~Heinemeyer, \emph{{A 96 GeV Higgs boson
  in the N2HDM}},
  \href{https://doi.org/10.1140/epjc/s10052-019-7561-2}{\emph{Eur. Phys. J. C}
  {\bfseries 80} (2020) 2} [\href{https://arxiv.org/abs/1903.11661}{{\ttfamily
  1903.11661}}].

\bibitem{Cline:2019okt}
J.M.~Cline and T.~Toma, \emph{{Pseudo-Goldstone dark matter confronts cosmic
  ray and collider anomalies}},
  \href{https://doi.org/10.1103/PhysRevD.100.035023}{\emph{Phys. Rev. D}
  {\bfseries 100} (2019) 035023}
  [\href{https://arxiv.org/abs/1906.02175}{{\ttfamily 1906.02175}}].

\bibitem{Cao:2019ofo}
J.~Cao, X.~Jia, Y.~Yue, H.~Zhou and P.~Zhu, \emph{{96 GeV diphoton excess in
  seesaw extensions of the natural NMSSM}},
  \href{https://doi.org/10.1103/PhysRevD.101.055008}{\emph{Phys. Rev. D}
  {\bfseries 101} (2020) 055008}
  [\href{https://arxiv.org/abs/1908.07206}{{\ttfamily 1908.07206}}].

\bibitem{Aguilar-Saavedra:2020wrj}
J.A.~Aguilar-Saavedra and F.R.~Joaquim, \emph{{Multiphoton signals of a (96
  GeV?) stealth boson}},
  \href{https://doi.org/10.1140/epjc/s10052-020-7952-4}{\emph{Eur. Phys. J. C}
  {\bfseries 80} (2020) 403}
  [\href{https://arxiv.org/abs/2002.07697}{{\ttfamily 2002.07697}}].

\bibitem{CMSnewtalk}
S.~Gascon-Shotkin, \emph{{{\rm CMS}, Talk at MoriondEW: Searches for additional
  Higgs bosons at low mass,
  \href{https://indico.in2p3.fr/event/29681/timetable/?view=standard_numbered\#84-searches-for-additional-hig}{\rm
  indico.cern.ch}}},  2023.

\bibitem{Heinemeyer:2018wzl}
S.~Heinemeyer and T.~Stefaniak, \emph{{A Higgs Boson at 96 GeV?!}},
  \href{https://doi.org/10.22323/1.339.0016}{\emph{PoS} {\bfseries CHARGED2018}
  (2019) 016} [\href{https://arxiv.org/abs/1812.05864}{{\ttfamily
  1812.05864}}].

\bibitem{Azatov:2012bz}
A.~Azatov, R.~Contino and J.~Galloway, \emph{{Model-Independent Bounds on a
  Light Higgs}}, \href{https://doi.org/10.1007/JHEP04(2012)127}{\emph{JHEP}
  {\bfseries 04} (2012) 127} [\href{https://arxiv.org/abs/1202.3415}{{\ttfamily
  1202.3415}}].

\bibitem{Heinemeyer:2018jcd}
S.~Heinemeyer, \emph{{A Higgs boson below 125 GeV?!}},
  \href{https://doi.org/10.1142/S0217751X18440062}{\emph{Int. J. Mod. Phys. A}
  {\bfseries 33} (2018) 1844006}.

\bibitem{Biekotter:2021ovi}
T.~Biek\"otter and M.O.~Olea-Romacho, \emph{{Reconciling Higgs physics and
  pseudo-Nambu-Goldstone dark matter in the S2HDM using a genetic algorithm}},
  \href{https://doi.org/10.1007/JHEP10(2021)215}{\emph{JHEP} {\bfseries 10}
  (2021) 215} [\href{https://arxiv.org/abs/2108.10864}{{\ttfamily
  2108.10864}}].

\bibitem{Biekotter:2021qbc}
T.~Biek\"otter, A.~Grohsjean, S.~Heinemeyer, C.~Schwanenberger and G.~Weiglein,
  \emph{{Possible indications for new Higgs bosons in the reach of the LHC:
  N2HDM and NMSSM interpretations}},
  \href{https://doi.org/10.1140/epjc/s10052-022-10099-1}{\emph{Eur. Phys. J. C}
  {\bfseries 82} (2022) 178}
  [\href{https://arxiv.org/abs/2109.01128}{{\ttfamily 2109.01128}}].

\bibitem{Heinemeyer:2021msz}
S.~Heinemeyer, C.~Li, F.~Lika, G.~Moortgat-Pick and S.~Paasch,
  \emph{{Phenomenology of a 96~GeV Higgs boson in the 2HDM with an additional
  singlet}}, \href{https://doi.org/10.1103/PhysRevD.106.075003}{\emph{Phys.
  Rev. D} {\bfseries 106} (2022) 075003}
  [\href{https://arxiv.org/abs/2112.11958}{{\ttfamily 2112.11958}}].

\bibitem{Biekotter:2022jyr}
T.~Biek\"otter, S.~Heinemeyer and G.~Weiglein, \emph{{Mounting evidence for a
  95 GeV Higgs boson}},
  \href{https://doi.org/10.1007/JHEP08(2022)201}{\emph{JHEP} {\bfseries 08}
  (2022) 201} [\href{https://arxiv.org/abs/2203.13180}{{\ttfamily
  2203.13180}}].

\bibitem{Biekotter:2022abc}
T.~Biek\"otter, S.~Heinemeyer and G.~Weiglein, \emph{{Excesses in the low-mass
  Higgs-boson search and the $W$-boson mass measurement}},
  \href{https://arxiv.org/abs/2204.05975}{{\ttfamily 2204.05975}}.

\bibitem{Jiang:2019soj}
X.-M.~Jiang, C.~Cai, Z.-H.~Yu, Y.-P.~Zeng and H.-H.~Zhang,
  \emph{{Pseudo-Nambu-Goldstone dark matter and two-Higgs-doublet models}},
  \href{https://doi.org/10.1103/PhysRevD.100.075011}{\emph{Phys. Rev. D}
  {\bfseries 100} (2019) 075011}
  [\href{https://arxiv.org/abs/1907.09684}{{\ttfamily 1907.09684}}].

\bibitem{Biekotter:2021ysx}
T.~Biek\"otter, S.~Heinemeyer, J.M.~No, M.O.~Olea and G.~Weiglein, \emph{{Fate
  of electroweak symmetry in the early Universe: Non-restoration and trapped
  vacua in the N2HDM}},
  \href{https://doi.org/10.1088/1475-7516/2021/06/018}{\emph{JCAP} {\bfseries
  06} (2021) 018} [\href{https://arxiv.org/abs/2103.12707}{{\ttfamily
  2103.12707}}].

\bibitem{Fayet:1976et}
P.~Fayet, \emph{{Supersymmetry and Weak, Electromagnetic and Strong
  Interactions}},
  \href{https://doi.org/10.1016/0370-2693(76)90319-1}{\emph{Phys. Lett. B}
  {\bfseries 64} (1976) 159}.

\bibitem{Mrazek:2011iu}
J.~Mrazek, A.~Pomarol, R.~Rattazzi, M.~Redi, J.~Serra and A.~Wulzer, \emph{{The
  Other Natural Two Higgs Doublet Model}},
  \href{https://doi.org/10.1016/j.nuclphysb.2011.07.008}{\emph{Nucl. Phys. B}
  {\bfseries 853} (2011) 1} [\href{https://arxiv.org/abs/1105.5403}{{\ttfamily
  1105.5403}}].

\bibitem{Kim:1986ax}
J.E.~Kim, \emph{{Light Pseudoscalars, Particle Physics and Cosmology}},
  \href{https://doi.org/10.1016/0370-1573(87)90017-2}{\emph{Phys. Rept.}
  {\bfseries 150} (1987) 1}.

\bibitem{Barger:2008jx}
V.~Barger, P.~Langacker, M.~McCaskey, M.~Ramsey-Musolf and G.~Shaughnessy,
  \emph{{Complex Singlet Extension of the Standard Model}},
  \href{https://doi.org/10.1103/PhysRevD.79.015018}{\emph{Phys. Rev. D}
  {\bfseries 79} (2009) 015018}
  [\href{https://arxiv.org/abs/0811.0393}{{\ttfamily 0811.0393}}].

\bibitem{PandaX-4T:2021bab}
{\scshape PandaX-4T} collaboration, \emph{{Dark Matter Search Results from the
  PandaX-4T Commissioning Run}},
  \href{https://doi.org/10.1103/PhysRevLett.127.261802}{\emph{Phys. Rev. Lett.}
  {\bfseries 127} (2021) 261802}
  [\href{https://arxiv.org/abs/2107.13438}{{\ttfamily 2107.13438}}].

\bibitem{XENON:2018voc}
{\scshape XENON} collaboration, \emph{{Dark Matter Search Results from a One
  Ton-Year Exposure of XENON1T}},
  \href{https://doi.org/10.1103/PhysRevLett.121.111302}{\emph{Phys. Rev. Lett.}
  {\bfseries 121} (2018) 111302}
  [\href{https://arxiv.org/abs/1805.12562}{{\ttfamily 1805.12562}}].

\bibitem{LZ:2022ufs}
{\scshape LZ} collaboration, \emph{{First Dark Matter Search Results from the
  LUX-ZEPLIN (LZ) Experiment}},
  \href{https://arxiv.org/abs/2207.03764}{{\ttfamily 2207.03764}}.

\bibitem{Branco:2011iw}
G.C.~Branco, P.M.~Ferreira, L.~Lavoura, M.N.~Rebelo, M.~Sher and J.P.~Silva,
  \emph{{Theory and phenomenology of two-Higgs-doublet models}},
  \href{https://doi.org/10.1016/j.physrep.2012.02.002}{\emph{Phys. Rept.}
  {\bfseries 516} (2012) 1} [\href{https://arxiv.org/abs/1106.0034}{{\ttfamily
  1106.0034}}].

\bibitem{Barbieri:2013nka}
R.~Barbieri, D.~Buttazzo, K.~Kannike, F.~Sala and A.~Tesi, \emph{{One or more
  Higgs bosons?}},
  \href{https://doi.org/10.1103/PhysRevD.88.055011}{\emph{Phys. Rev. D}
  {\bfseries 88} (2013) 055011}
  [\href{https://arxiv.org/abs/1307.4937}{{\ttfamily 1307.4937}}].

\bibitem{Bechtle:2008jh}
P.~Bechtle, O.~Brein, S.~Heinemeyer, G.~Weiglein and K.E.~Williams,
  \emph{{HiggsBounds: Confronting Arbitrary Higgs Sectors with Exclusion Bounds
  from LEP and the Tevatron}},
  \href{https://doi.org/10.1016/j.cpc.2009.09.003}{\emph{Comput. Phys. Commun.}
  {\bfseries 181} (2010) 138}
  [\href{https://arxiv.org/abs/0811.4169}{{\ttfamily 0811.4169}}].

\bibitem{Bechtle:2011sb}
P.~Bechtle, O.~Brein, S.~Heinemeyer, G.~Weiglein and K.E.~Williams,
  \emph{{HiggsBounds 2.0.0: Confronting Neutral and Charged Higgs Sector
  Predictions with Exclusion Bounds from LEP and the Tevatron}},
  \href{https://doi.org/10.1016/j.cpc.2011.07.015}{\emph{Comput. Phys. Commun.}
  {\bfseries 182} (2011) 2605}
  [\href{https://arxiv.org/abs/1102.1898}{{\ttfamily 1102.1898}}].

\bibitem{Bechtle:2013wla}
P.~Bechtle, O.~Brein, S.~Heinemeyer, O.~St\r{a}l, T.~Stefaniak, G.~Weiglein
  et~al., \emph{{$\mathsf{HiggsBounds}-4$: Improved Tests of Extended Higgs
  Sectors against Exclusion Bounds from LEP, the Tevatron and the LHC}},
  \href{https://doi.org/10.1140/epjc/s10052-013-2693-2}{\emph{Eur. Phys. J. C}
  {\bfseries 74} (2014) 2693}
  [\href{https://arxiv.org/abs/1311.0055}{{\ttfamily 1311.0055}}].

\bibitem{Bechtle:2020pkv}
P.~Bechtle, D.~Dercks, S.~Heinemeyer, T.~Klingl, T.~Stefaniak, G.~Weiglein
  et~al., \emph{{HiggsBounds-5: Testing Higgs Sectors in the LHC 13 TeV Era}},
  \href{https://doi.org/10.1140/epjc/s10052-020-08557-9}{\emph{Eur. Phys. J. C}
  {\bfseries 80} (2020) 1211}
  [\href{https://arxiv.org/abs/2006.06007}{{\ttfamily 2006.06007}}].

\bibitem{Bahl:2022igd}
H.~Bahl, T.~Biek\"otter, S.~Heinemeyer, C.~Li, S.~Paasch, G.~Weiglein et~al.,
  \emph{{HiggsTools: BSM scalar phenomenology with new versions of HiggsBounds
  and HiggsSignals}},  \href{https://arxiv.org/abs/2210.09332}{{\ttfamily
  2210.09332}}.

\bibitem{Bechtle:2013xfa}
P.~Bechtle, S.~Heinemeyer, O.~St\r{a}l, T.~Stefaniak and G.~Weiglein,
  \emph{{$HiggsSignals$: Confronting arbitrary Higgs sectors with measurements
  at the Tevatron and the LHC}},
  \href{https://doi.org/10.1140/epjc/s10052-013-2711-4}{\emph{Eur. Phys. J. C}
  {\bfseries 74} (2014) 2711}
  [\href{https://arxiv.org/abs/1305.1933}{{\ttfamily 1305.1933}}].

\bibitem{Bechtle:2014ewa}
P.~Bechtle, S.~Heinemeyer, O.~St\r{a}l, T.~Stefaniak and G.~Weiglein,
  \emph{{Probing the Standard Model with Higgs signal rates from the Tevatron,
  the LHC and a future ILC}},
  \href{https://doi.org/10.1007/JHEP11(2014)039}{\emph{JHEP} {\bfseries 11}
  (2014) 039} [\href{https://arxiv.org/abs/1403.1582}{{\ttfamily 1403.1582}}].

\bibitem{Bechtle:2020uwn}
P.~Bechtle, S.~Heinemeyer, T.~Klingl, T.~Stefaniak, G.~Weiglein and
  J.~Wittbrodt, \emph{{HiggsSignals-2: Probing new physics with precision Higgs
  measurements in the LHC 13 TeV era}},
  \href{https://doi.org/10.1140/epjc/s10052-021-08942-y}{\emph{Eur. Phys. J. C}
  {\bfseries 81} (2021) 145}
  [\href{https://arxiv.org/abs/2012.09197}{{\ttfamily 2012.09197}}].

\bibitem{Muhlleitner:2016mzt}
M.~Muhlleitner, M.O.P.~Sampaio, R.~Santos and J.~Wittbrodt, \emph{{The N2HDM
  under Theoretical and Experimental Scrutiny}},
  \href{https://doi.org/10.1007/JHEP03(2017)094}{\emph{JHEP} {\bfseries 03}
  (2017) 094} [\href{https://arxiv.org/abs/1612.01309}{{\ttfamily
  1612.01309}}].

\bibitem{Engeln:2018mbg}
I.~Engeln, M.~M\"uhlleitner and J.~Wittbrodt, \emph{{N2HDECAY: Higgs Boson
  Decays in the Different Phases of the N2HDM}},
  \href{https://doi.org/10.1016/j.cpc.2018.07.020}{\emph{Comput. Phys. Commun.}
  {\bfseries 234} (2019) 256}
  [\href{https://arxiv.org/abs/1805.00966}{{\ttfamily 1805.00966}}].

\bibitem{Haller:2018nnx}
J.~Haller, A.~Hoecker, R.~Kogler, K.~M\"onig, T.~Peiffer and J.~Stelzer,
  \emph{{Update of the global electroweak fit and constraints on
  two-Higgs-doublet models}},
  \href{https://doi.org/10.1140/epjc/s10052-018-6131-3}{\emph{Eur. Phys. J. C}
  {\bfseries 78} (2018) 675}
  [\href{https://arxiv.org/abs/1803.01853}{{\ttfamily 1803.01853}}].

\bibitem{Grimus:2008nb}
W.~Grimus, L.~Lavoura, O.M.~Ogreid and P.~Osland, \emph{{The Oblique parameters
  in multi-Higgs-doublet models}},
  \href{https://doi.org/10.1016/j.nuclphysb.2008.04.019}{\emph{Nucl. Phys. B}
  {\bfseries 801} (2008) 81} [\href{https://arxiv.org/abs/0802.4353}{{\ttfamily
  0802.4353}}].

\bibitem{CDF:2022hxs}
{\scshape CDF} collaboration, \emph{{High-precision measurement of the $W$
  boson mass with the CDF II detector}},
  \href{https://doi.org/10.1126/science.abk1781}{\emph{Science} {\bfseries 376}
  (2022) 170}.

\bibitem{Planck:2018vyg}
{\scshape Planck} collaboration, \emph{{Planck 2018 results. VI. Cosmological
  parameters}},
  \href{https://doi.org/10.1051/0004-6361/201833910}{\emph{Astron. Astrophys.}
  {\bfseries 641} (2020) A6}
  [\href{https://arxiv.org/abs/1807.06209}{{\ttfamily 1807.06209}}].

\bibitem{Belanger:2018ccd}
G.~B\'elanger, F.~Boudjema, A.~Goudelis, A.~Pukhov and B.~Zaldivar,
  \emph{{micrOMEGAs5.0 : Freeze-in}},
  \href{https://doi.org/10.1016/j.cpc.2018.04.027}{\emph{Comput. Phys. Commun.}
  {\bfseries 231} (2018) 173}
  [\href{https://arxiv.org/abs/1801.03509}{{\ttfamily 1801.03509}}].

\bibitem{Biekotter:2022bxp}
T.~Biek\"otter, P.~Gabriel, M.O.~Olea-Romacho and R.~Santos, \emph{{Direct
  detection of pseudo-Nambu-Goldstone dark matter in a two Higgs doublet plus
  singlet extension of the SM}},
  \href{https://doi.org/10.1007/JHEP10(2022)126}{\emph{JHEP} {\bfseries 10}
  (2022) 126} [\href{https://arxiv.org/abs/2207.04973}{{\ttfamily
  2207.04973}}].

\bibitem{Gross:2017dan}
C.~Gross, O.~Lebedev and T.~Toma, \emph{{Cancellation Mechanism for
  Dark-Matter\textendash{}Nucleon Interaction}},
  \href{https://doi.org/10.1103/PhysRevLett.119.191801}{\emph{Phys. Rev. Lett.}
  {\bfseries 119} (2017) 191801}
  [\href{https://arxiv.org/abs/1708.02253}{{\ttfamily 1708.02253}}].

\bibitem{Abe:2020dut}
Y.~Abe, Y.~Hamada, T.~Ohata, K.~Suzuki and K.~Yoshioka, \emph{{TeV-scale
  Majorogenesis}}, \href{https://doi.org/10.1007/JHEP07(2020)105}{\emph{JHEP}
  {\bfseries 07} (2020) 105}
  [\href{https://arxiv.org/abs/2004.00599}{{\ttfamily 2004.00599}}].

\bibitem{Abe:2020iph}
Y.~Abe, T.~Toma and K.~Tsumura, \emph{{Pseudo-Nambu-Goldstone dark matter from
  gauged $U(1)_{B-L}$ symmetry}},
  \href{https://doi.org/10.1007/JHEP05(2020)057}{\emph{JHEP} {\bfseries 05}
  (2020) 057} [\href{https://arxiv.org/abs/2001.03954}{{\ttfamily
  2001.03954}}].

\bibitem{Okada:2020zxo}
N.~Okada, D.~Raut and Q.~Shafi, \emph{{Pseudo-Goldstone dark matter in a gauged
  $B-L$ extended standard model}},
  \href{https://doi.org/10.1103/PhysRevD.103.055024}{\emph{Phys. Rev. D}
  {\bfseries 103} (2021) 055024}
  [\href{https://arxiv.org/abs/2001.05910}{{\ttfamily 2001.05910}}].

\bibitem{Mohapatra:2023aei}
R.N.~Mohapatra and N.~Okada, \emph{{Conformal B-L and pseudo-Goldstone dark
  matter}}, \href{https://doi.org/10.1103/PhysRevD.107.095023}{\emph{Phys. Rev.
  D} {\bfseries 107} (2023) 095023}
  [\href{https://arxiv.org/abs/2302.11072}{{\ttfamily 2302.11072}}].

\bibitem{Liu:2022evb}
D.-Y.~Liu, C.~Cai, X.-M.~Jiang, Z.-H.~Yu and H.-H.~Zhang, \emph{{Ultraviolet
  completion of pseudo-Nambu-Goldstone dark matter with a hidden U(1) gauge
  symmetry}}, \href{https://doi.org/10.1007/JHEP02(2023)104}{\emph{JHEP}
  {\bfseries 02} (2023) 104}
  [\href{https://arxiv.org/abs/2208.06653}{{\ttfamily 2208.06653}}].

\bibitem{DEAP_JMLR2012}
F.-A.~Fortin, F.-M.~{De Rainville}, M.-A.~Gardner, M.~Parizeau and C.~Gagn\'e,
  \emph{{DEAP}: Evolutionary algorithms made easy}, {\emph{Journal of Machine
  Learning Research} {\bfseries 13} (2012) 2171}.

\bibitem{CMS-PAS-EXO-21-018}
{\scshape CMS} collaboration, \emph{{Search for dilepton resonances from decays
  of (pseudo)scalar bosons produced in association with a massive vector boson
  or top quark anti-top quark pair at $\sqrt{s}=13~\mathrm{TeV}$}},  Tech. Rep.
  \href{https://cds.cern.ch/record/2815307}{CMS-PAS-EXO-21-018} (2022).

\bibitem{Iguro:2022dok}
S.~Iguro, T.~Kitahara and Y.~Omura, \emph{{Scrutinizing the 95\textendash{}100
  GeV di-tau excess in the top associated process}},
  \href{https://doi.org/10.1140/epjc/s10052-022-11028-y}{\emph{Eur. Phys. J. C}
  {\bfseries 82} (2022) 1053}
  [\href{https://arxiv.org/abs/2205.03187}{{\ttfamily 2205.03187}}].

\bibitem{ATLAS:2022yrq}
{\scshape ATLAS} collaboration, \emph{{Measurements of Higgs boson production
  cross-sections in the~$H\to\tau^{+}\tau^{-}$ decay channel in pp collisions
  at $ \sqrt{s} $ = 13 TeV with the ATLAS detector}},
  \href{https://doi.org/10.1007/JHEP08(2022)175}{\emph{JHEP} {\bfseries 08}
  (2022) 175} [\href{https://arxiv.org/abs/2201.08269}{{\ttfamily
  2201.08269}}].

\bibitem{Coloretti:2023wng}
G.~Coloretti, A.~Crivellin, S.~Bhattacharya and B.~Mellado, \emph{{Searching
  for Low-Mass Resonances Decaying into $W$ Bosons}},
  \href{https://arxiv.org/abs/2302.07276}{{\ttfamily 2302.07276}}.

\bibitem{ATLAS:2022ooq}
{\scshape ATLAS} collaboration, \emph{{Measurements of Higgs boson production
  by gluon$-$gluon fusion and vector-boson fusion using $H\rightarrow W W^*
  \rightarrow e\nu \mu\nu$ decays in $pp$ collisions at $\sqrt{s}=13$ TeV with
  the ATLAS detector}},  \href{https://arxiv.org/abs/2207.00338}{{\ttfamily
  2207.00338}}.

\bibitem{CMS:2022uhn}
{\scshape CMS} collaboration, \emph{{Measurements of the Higgs boson production
  cross section and couplings in the W boson pair decay channel in
  proton-proton collisions at $\sqrt{s}$ = 13 TeV}},
  \href{https://arxiv.org/abs/2206.09466}{{\ttfamily 2206.09466}}.

\bibitem{Hollik:2018yek}
W.G.~Hollik, S.~Liebler, G.~Moortgat-Pick, S.~Pa\ss{}ehr and G.~Weiglein,
  \emph{{Phenomenology of the inflation-inspired NMSSM at the electroweak
  scale}}, \href{https://doi.org/10.1140/epjc/s10052-019-6561-6}{\emph{Eur.
  Phys. J. C} {\bfseries 79} (2019) 75}
  [\href{https://arxiv.org/abs/1809.07371}{{\ttfamily 1809.07371}}].

\bibitem{Biekotter:2019gtq}
T.~Biek\"otter, S.~Heinemeyer and C.~Mu\~noz, \emph{{Precise prediction for the
  Higgs-Boson masses in the $\mu\nu$SSM with three right-handed neutrino
  superfields}},
  \href{https://doi.org/10.1140/epjc/s10052-019-7175-8}{\emph{Eur. Phys. J. C}
  {\bfseries 79} (2019) 667}
  [\href{https://arxiv.org/abs/1906.06173}{{\ttfamily 1906.06173}}].

\bibitem{Choi:2019yrv}
K.~Choi, S.H.~Im, K.S.~Jeong and C.B.~Park, \emph{{Light Higgs bosons in the
  general NMSSM}},
  \href{https://doi.org/10.1140/epjc/s10052-019-7473-1}{\emph{Eur. Phys. J. C}
  {\bfseries 79} (2019) 956}
  [\href{https://arxiv.org/abs/1906.03389}{{\ttfamily 1906.03389}}].

\bibitem{Cepeda:2019klc}
M.~Cepeda et~al., \emph{{Report from Working Group 2}: {Higgs Physics at the
  HL-LHC and HE-LHC}},
  \href{https://doi.org/10.23731/CYRM-2019-007.221}{\emph{CERN Yellow Rep.
  Monogr.} {\bfseries 7} (2019) 221}
  [\href{https://arxiv.org/abs/1902.00134}{{\ttfamily 1902.00134}}].

\bibitem{Bambade:2019fyw}
P.~Bambade et~al., \emph{{The International Linear Collider: A Global
  Project}},  \href{https://arxiv.org/abs/1903.01629}{{\ttfamily 1903.01629}}.

\bibitem{deBlas:2019rxi}
J.~de~Blas et~al., \emph{{Higgs Boson Studies at Future Particle Colliders}},
  \href{https://doi.org/10.1007/JHEP01(2020)139}{\emph{JHEP} {\bfseries 01}
  (2020) 139} [\href{https://arxiv.org/abs/1905.03764}{{\ttfamily
  1905.03764}}].

\bibitem{Drechsel:2018mgd}
P.~Drechsel, G.~Moortgat-Pick and G.~Weiglein, \emph{{Prospects for direct
  searches for light Higgs bosons at the ILC with 250 GeV}},
  \href{https://doi.org/10.1140/epjc/s10052-020-08438-1}{\emph{Eur. Phys. J. C}
  {\bfseries 80} (2020) 922}
  [\href{https://arxiv.org/abs/1801.09662}{{\ttfamily 1801.09662}}].

\bibitem{Wang:2020lkq}
Y.~Wang, M.~Berggren and J.~List, \emph{{ILD Benchmark: Search for Extra
  Scalars Produced in Association with a $Z$ boson at $\sqrt{s}=500$ GeV}},
  \href{https://arxiv.org/abs/2005.06265}{{\ttfamily 2005.06265}}.

\end{thebibliography}\endgroup

\end{document}